\newcommand{\eps}{\varepsilon}
\newcommand{\Ham}{{\cal{H}}}
\newcommand{\Ope}{{\cal{O}}}
\newcommand{\M}{{\bf M}}
\newcommand{\I}{{\bf I}}
\newcommand{\J}{{\bf J}}

\newcommand{\Exp}[1]{{\it e}^{#1}}
\newcommand{\Erw}[1]{\langle #1\rangle}
\newcommand{\Bra}[1]{\langle #1\mid}
\newcommand{\Ket}[1]{\mid #1\rangle}
\newcommand{\abs}[1]{\mid #1\mid}

\documentstyle[aps,epsfig,prb,eqsecnum]{revtex}

\begin{document}
\draft
\twocolumn[\hsize\textwidth\columnwidth\hsize
          \csname @twocolumnfalse\endcsname
\title{
{\vspace{-7mm}\small\framebox[14cm][l]{%
submitted an PRB, 10.11.1999, } \\
\vspace{2mm}}%
Characterization of localized hole states in 
Pr$_{1+x}$Ba$_{2-x}$Cu$_3$O$_{6+y}$ by nuclear magnetic resonance}
\author{M. W. Pieper}
\address{TU-Wien, Wiedner Hauptstr. 8, A-1040 Wien, Austria, email: Pieper@xphys.tuwien.ac.at}
\author{F. Wiekhorst}
\address{Universit\"at Hamburg, Jungiusstr.11, D-20355 Hamburg}
\author{T. Wolf}
\address{Forschungszentrum Karlsruhe, Institut f. Technische Physik}
\date{\today}
\maketitle

\begin{abstract}
  We investigated the charge distribution in
  Pr$_{1+x}$Ba$_{2-x}$Cu$_3$O$_{6+y}$-crystals at liquid-He
  temperature by means of $^{63,65}$Cu- and $^{141}$Pr-NMR and NQR.
  The electric field gradients determined for the different oxygen
  coordinations of Cu(1) on the chains confirm the model that the hole
  states are localized in the O$2p\pi$-orbitals of the CuO$_2$-planes.
  The intensity and line-shape analysis of the Cu(1)- and
  Pr-resonances in the oxygen doping series ($x\approx 0, y=0\ldots
  1$) and in the Pr/Ba solid-solution series (at $y=1$) allows us to
  assign the Pr-resonance to Pr at RE-sites. Therefore, we can
  investigate the charge distribution in the CuO$_2$-Pr-CuO$_2$-layers
  by crystal field analysis of the Pr-signal. To consistently describe
  our results with neutron scattering data we propose that two
  Pr-states are present in the RE-layer, and ascribe them to Pr with
  and one without a hole localized in the oxygen coordination shell.
  NMR detects only the former, with a virtually nonmagnetic singlet
  ground state, while space-integral techniques are dominated by the
  latter, due to a quasi-doublet ground state and antiferromagnetism of
  its Pr-moments at low temperature.
\end{abstract}

\pacs{74.72.-h, 76.60.-k, 74.25.Jb, 75.50.Ee}
]

\narrowtext

\section{Introduction}

PrBa$_2$Cu$_3$O$_7$ is the one insulating member of the isostructural
series REBa$_2$Cu$_3$O$_7$ of high-T$_c$
superconductors\cite{rad216}  (RE=Rare Earth, Y, La, abbreviated
below RE123). Simultaneously the magnetic properties of Pr differ
dramatically from the reasonably well understood $4f-$magnetism of all
other rare earths in this structure. This extraordinary behaviour of
Pr attracted considerable interest from experimental and theoretical
physics throughout the last decade because it might shed some light on
the microscopic origin of high-T$_c$ superconductivity in the cuprates.
From the point of view of applied physics it is important to understand
the physical properties of the material because of its potential use
in superconductor-insulator heterostructures.

There is a general consensus that the structural key element of the
cuprate superconductors are the CuO$_2$-planes. The undoped
CuO$_2$-plane is an antiferromagnetic charge transfer insulator:
Photoemission spectroscopy shows that the on-site correlation energy
of the Cu-$d_{x^2-y^2}$ electrons is large enough to push the lower
Hubbard band below the 2p-oxygen band\cite{Huefner}. Upon hole doping
an insulator-metal transition (MIT) takes place, with a
superconducting ground state in a certain range of hole
concentrations. The dependence of the superconducting transition
temperature $T_c$ on the hole concentration $n_h$ in the CuO$_2$-plane
follows a universal law close to $T_c/T_{c
  max}=1-((n_h-n_{opt})/\Delta_n)^2$, where this has positive
solutions\cite{pic333,tal438} (note that other dependencies have been
proposed, e.g. the trapezoid shape in ref.\cite{zha439}). In contrast to
the large differences in the maximum transition temperature $T_{c
  max}$ between cuprates, the variations of the optimum hole
concentration $n_{opt}\approx 0.3/$CuO$_2$-unit and the threshold
concentration $\Delta_n$ for superconductivity appear to be rather
small. The fact that Pr123 is an antiferromagnetic insulator instead
of metallic and superconducting like the other members of the RE123
series may then be due to either a suppression of the doping, or a
detail of the electronic structure lowering $T_{c max}$ dramatically.

There is no obvious and generally accepted parameter connected with
$T_{c max}$ of the cuprates. In RE123 one finds nearly independent of
the RE $T_{c max}=93\pm 4$~K. Magnetic pair breaking by the Pr-$4f$
moments, enhanced by a hybridization with the bands in the
CuO$_2$-planes, has been invoked as a mechanism to suppress
superconductivity by lowering $T_{c max}$. A phenomenological
combination of Abrikosov-Gorkov pair breaking and hole depletion can
successfully describe the dependence of the superconducting $T_c$ on
the Pr-concentration in Pr$_z$RE$_{1-z}$123,\cite{neu416,cao420} and
it has been argued that the exceptionally high ordering temperature of
the Pr-sublattice, $T_N=12\ldots 18$~K compared to 2.1~K for Gd, also
supports this view.\cite{ska331} However, the results cannot be
described by pair breaking alone, it is necessary to take hole
depletion into account. Otherwise Pr would not suppress
superconductivity and the MIT at the same critical Pr-concentration,
and $T_c$ would drop linearly at small $x$, which is not
observed.\cite{cao420,neu416,tom422} Furthermore, there is evidence
against pair breaking,\cite{fra428} and it is possible to model
$T_c(z)$ without magnetic pair breaking under the assumption that the
influence of Pr on the doping depends on its
concentration.\cite{tom422,wan423} Even if one agrees that pair
breaking is indicated experimentally at least near the critical
Pr-concentration, it is not obvious that a Pr-moment is involved since
near the MIT localized Cu-moments in the CuO$_2$-layers may be
candidates for pair breakers as well.

Therefore, the current attempts to explain the role of Pr focus on the
suppression of the hole doping. There is clear experimental evidence
that Pr does not simply fill the hole states with its $4f$-electrons.
Various techniques,\cite{uma318,fis413,lun417} including inelastic neutron
scattering (INS)\cite{hil147,boo205} and our own previous NMR
work,\cite{neh199,neh352} show that the valence of Pr is close to
three, the same as for the other RE, and stays constant independent of
the oxygen content $x$.\cite{lop447}  Furthermore, spectroscopic methods
sensitive to the density of O-$2p$ hole states detect similar
amounts in Pr123 as in the other RE123.\cite{fin200,har410}

The by now generally accepted model for the suppression of hole doping
was proposed by Fehrenbacher and Rice. In their model the holes become
localized in the antibonding O-$2p\pi$-orbitals of the
planes.\cite{feh198} These states are favourable to the
$2p\sigma$-states populated in the absence of Pr due to their
hybridization with the $4f^2$-shell of Pr. This view has been
supported and extended by band structure calculations of Liechtenstein
and Mazin, showing that with increasing radius of the RE just for Pr a
new band with $4f$-character crosses the Fermi
level,\cite{lie197,maz357} and by Wang et al., who explicitly
included a third band from the chain layers to describe the hole
concentration as a function of the Pr-concentration in the
RE-layer.\cite{wan423} As an alternative to the hybridization models it
has been discussed that Pr might push the hole states in a band
associated with the apex oxygen.\cite{kho000,dre440} 

Experimentally, the hybridization shows (among other techniques) in
detailed Pr-O bond length analysis,\cite{gui232} in the strong
broadening of the crystal field transitions in
INS,\cite{hil147,boo205} and x-ray absorption.\cite{hu431,che414} Using
polarized x-rays Merz et al. were recently able to show that the
orientation of the 2p-hole states in the structure is indeed different
in Y123 and Pr123, giving very strong support to the hybridization
model.\cite{mer358} Clearly, a microscopic characterization of the
localized hole states is very desirable in order to distinguish
between the models and address open problems such as the mechanism of
the carrier localization or of the role of the Pr- and Cu-magnetism at
low temperature.

In RE123, doping of the CuO$_2$-planes is achieved by variation of the
oxygen content of the chain layer, and one of the basic ingredients of
hybridization models is that this doping mechanism is independent of
the RE-ion. A model for the dependence of the hole count $n_h$ on the
oxygen concentration in the chains $y$ has to address two problems,
namely the distribution of oxygen in the chains, and the hole count
induced by a given oxygen configuration.\cite{mcc434} The non-random
oxygen distribution in the chain layer is best described by the
asymmetric next-nearest neighbor Ising model.\cite{liu435,ali437} It
describes the various superstructures observed at intermediate $y$
(most important the formation of chains in Y123 at $y=0.5$) which
depend strongly on the RE-ion. L\"utgemeier et al. have shown that
Cu(1)-(chain-)site NQR is a suitable method to investigate this
distribution,\cite{hei3,lue329} but they did not include Pr123 in
their studies, so we analyse this case below in some detail.

The dependence of the hole count on the oxygen configuration with the
chain length as the most prominent parameter has been subject to
detailed theoretical studies in Y123.\cite{uim436,dre440} On the
experimental side the electric field gradient (EFG) at the Cu(1)-sites
is a sensitive probe of the charge balance between the planes and the
chain layer.\cite{amb429} Ohno et al. used this to estimate the number
of mobile carriers in the planes of metallic, fully oxidized
Pr$_z$Y$_{1-z}$123.\cite{ohn339} Below we use the Cu(1)-EFG at various
oxygen contents to verify experimentally the assumption underlying the
FR model, that the CuO-chains dope holes into the
CuO$_2$-RE-CuO$_2$-layers, independent of the presence of Pr.

One might expect that the special arrangement of charges in the
hybridized state proposed by Fehrenbacher and Rice shows in the
crystal field parameters of the $4f$-shell of Pr. Unfortunately, the
lines from crystal field transitions in INS are severly broadened,
giving rise to large uncertainties in the symmetry as well as in the
energies of the low lying crystal field levels.\cite{hil147,boo205}
NMR is very sensitive especially to the symmetry of the ground state
and the position of the lowest excitations.\cite{bak441} In a previous
work we reported the first measurements of the $^{141}$Pr-resonance in
Pr123 for $y=1$.\cite{neh352} Most remarkably, we found a different
ground state symmetry for Pr$^{3+}$ from the one used to describe the
INS spectra. Moreover, the virtually nonmagnetic state of Pr at low
temperatures which is without doubt what is observed in our
NMR-measurements is in clear contrast to the antiferromagnetic
Pr-order detected by neutron diffraction and the homogeneous
suzeptibility.\cite{li52,boo306,ska331,uma419}

Our interpretation of the NMR-data has been strongly questioned
because of these findings,\cite{sta461,boo306,ska331} but no
consistent explanation for both, the NMR {\em and} the neutron data
has been proposed up to now. Therefore we extended the investigation
of the Pr-resonance to crystals over the whole range $y=0\ldots 1$ of
oxygen concentrations, and to crystals with different concentrations
in the Ba/Pr solid-solution system. On this basis we confirm our
initial assignment of the Pr-resonance to Pr on regular RE-sites. In
view of the strong doubts our brief description of the $4f-$ground
state in ref.\cite{neh352} did encounter we give below a rather
detailed discussion of our determination of the crystal field ground
states of Pr in this structure. This leads us to a consistent
description of the NMR and neutron experiments, basically by assuming
that the interpretation of both has been correct, and assigning two
different electronic states of Pr to the conflicting experimental
observations.

Finally we discuss the impact of this model on the understanding of
the low temperature magnetism of Pr and the other rare earths in this
remarkable material. As indicated above, the Pr-magnetism with its
Neel-temperature up to 18~K in Pr123 and in related
cuprates\cite{coo402,li52,sun445,hsi444,lai442} is as outstanding as
the suppression of superconductivity. It is tempting to relate the two
effects by the argument that the $4f-2p\pi-$hybridization suppresses
superconductivity by switching on pair breaking at Pr-moments, and
simultaneously induces the high $T_N$ via a substantial enhancement of
the $4f-4f$-exchange.\cite{ska331} We would like to caution about this
argument: While there is ample evidence for the hybridization itself
and we present more below, the contribution of pair breaking to the
suppression of superconductivity is much less clear (see above), and
the role of the $4f-$exchange in Pr-magnetic order seems even more
open. In fact the exchange is known to be small for all other rare
earths, the ordering of the other $4f-$moments is mostly due to
dipolar interactions.\cite{dro304} This is most convincingly
demonstrated by the small dependence of $T_N$ on magnetic dilution in
the rare earth sublattice for all magnetic
RE-ions.\cite{gua351,das403,keb226} All magnetic RE, including the
exceptional singlet ground-state Pr, order well above the 40\%
percolation threshold for dilution of the square lattice with nearest
neighbor exchange. The order is, therefore, ascribed to the long range
dipolar coupling of the magnetic rare earths, which is clearly not an
option in the case of Pr123.  The dependence of $T_N$ on oxygen
content is also puzzling, if hybridization enhanced exchange is the
origin. At $y=0$, where there is no hybridization of hole states,
$T_N$ is still as high as 12~K. In Pr123, $T_N$ increases with the
hole count, while it decreases for Nd in Nd123.\cite{dro304} In the
concluding section of the discussion we show that the increase of
$T_N$ of Pr is even more puzzling in view of our NMR-data, and discuss
induced magnetism of the Pr-sublattice as an alternative model.

\section{Experimental Details and Results}

\subsection{Sample characterization}

The main diffculties in the experimental verification of the models
arise from the complex doping mechanism in the RE123 series on the
hand, and from the strong influence of crystal preparation on the
material properties on the other. A remarkable example is the
preparation of crystals under special conditions that do become
superconducting, with $T_{c max}$ even higher than in the other RE123
($>100$~K under pressure).\cite{zou353,lus509} In order to provide a reliable
basis for comparison of results obtained with other samples we present
in this subsection a rather detailed description of the sample
preparation.

High quality Pr123 crystals of a size suitable for NMR ($\ge 10$ mg)
are still very difficult to prepare. One major
problem is the contamination of the Cu(1)-sublattice with Al from
Al$_2$O$_3$ crucibles (up to 30\%), because this has a strong
influence on the oxygen distribution\cite{wid415} as well
as on the magnetic structure of the Cu(2)-sublattice at low
temperature.\cite{bre350} We avoid the Al-contamination by use of
MgO-crucibles, which results in a much smaller contamination by Mg
($\approx 1\%$) because of the higher melting point of MgO.\cite{tag453}
The best stoichiometric crystals available are grown in
BaZrO$_3$-crucibles, but they have been too small for NMR up to now. 

\begin{figure}
\psfig{file=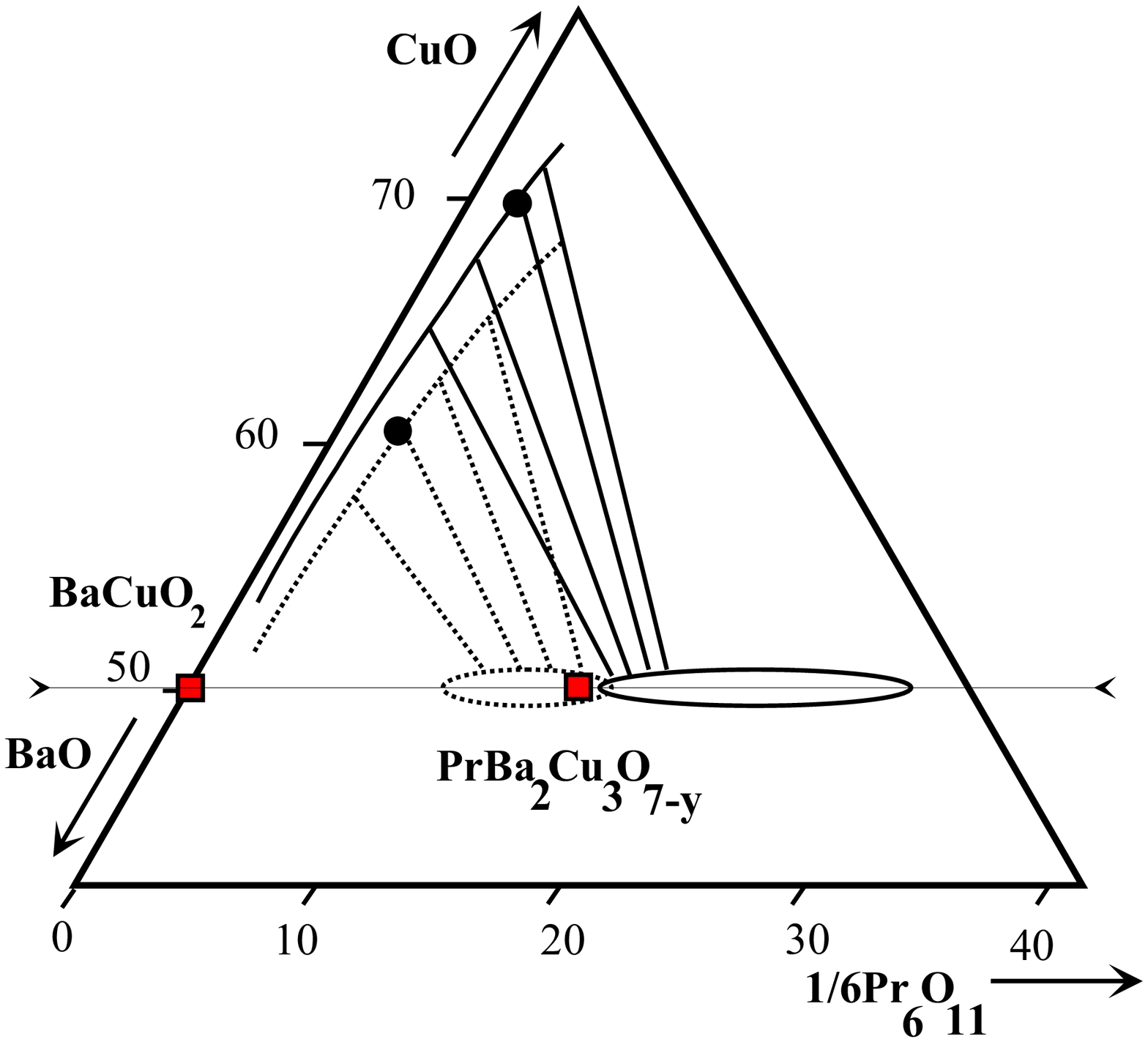,width=8cm,clip=}
\psfig{file=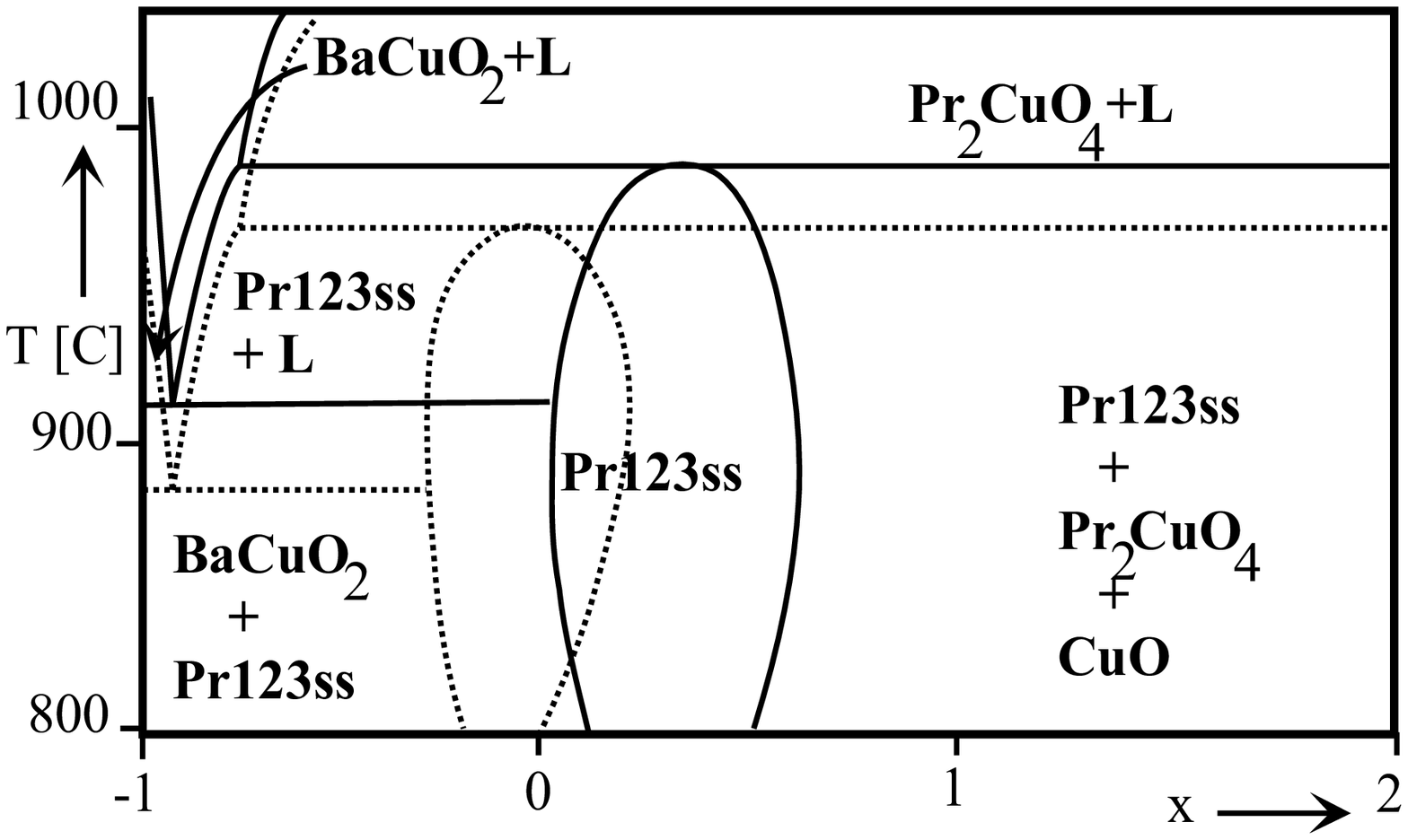,width=8cm,clip=}
\caption{Top: Schematic phase diagram of the Pr-Ba-Cu-O system at
  940$^\circ$C in 1 bar O$_2$ (solid lines) and in 63 mbar O$_2$
  (dotted lines). Stoichiometric phases are indicated by filled
  squares. A few conodal lines connecting the liquidus surface with
  the Pr/Ba solid solution system (Pr123ss) are shown with the
  starting compositions for the crystal growth marked by the dots.
  Bottom: Schematic phase diagram along the horizontal line
  corresponding to Pr$_{1+x}$Ba$_{2-x}$Cu$_3$O$_{7-y}$ in the upper
  part, again for 1 bar O$_2$ (solid), and 63 mbar O$_2$ (or 300 mbar
  air, dotted).}
\label{Pr123ss}
\end{figure}

A second problem is the existence of a finite solid solution range for
the RE/Ba-sublattices for the light rare earths (Nd, Pr,
La).\cite{par412,tag330,lin463,ber510} The solid solution range is a
four-dimensional volume in the phase diagram spanned by temperature
$T$, partial oxygen pressure $p(O_2)$, and the (2D-) triangle of
cation concentrations, sketched for fixed $T$ and $p(O_2)$ in
fig.\ref{Pr123ss}. The stable compositions form the surface of this
volume in the phase diagram with the ellipses in fig.\ref{Pr123ss}
(top) indicating cross sections. Also shown are the liquidus surfaces
and the conodal lines connecting them to the solid solution systems at
two different $p(O_2$). In the absence of other metals which may
substitute for Cu (e.g.  Al from the crucibles) one may assume a
constant Cu content. In this case the ellipses in fig.1 degenerate to
horizontal lines at the concentrations
R$_{1+x}$Ba$_{2-x}$Cu$_3$O$_{6+y}$, and the solid solution range has
only the three degrees of freedom $T$, $p(O_2)$, and $x$.

The accessible range of $x$ depends on temperature, oxygen partial
pressure, and on the rare earth radius. In general, the RE-content $x$
increases with increasing oxygen partial pressure and rare earth
radius, and with decreasing temperature, at least in the vicinity
below the peritectic temperature. We grew Pr123 crystals by the slow
cooling method.\cite{gol000} The Pr/Ba ratio of the growing crystals
was set by the Cu/Ba ratio of the flux via the conodal line to the
solidus surface. Temperature and oxygen pressure for variation of $x$
were chosen as marked by the dots in the phase diagram to obtain
crystals with high and low Pr-content along the conodal lines
indicated in the figure (see also tab.\ref{prepcond}).

\begin{table}
\label{tab1}
\begin{tabular}{llll}
 & $x<0$ & $x=0$ & $x>0$ \\
\hline
at\%Pr & 2      & 2         & 2         \\
at\%Ba & 37     & 30        & 28        \\
at\%Cu & 61     & 68        & 70        \\
\hline
 & 300 mbar air & 1 bar air & 1 bar O$_2$ \\
\hline
$[^\circ$C$]$& 970$\rightarrow$906& 1000$\rightarrow$939& 1000$\rightarrow$948 \\
$[^\circ$C/h$]$ & -0.4          & -0.5           & -0.35          \\
              & quench        & slow cool      & quench         \\
\hline
       & orthor.   & orthor.   & tetr.  \\
(100)/ & (010)     & (010)     & (111) \\
\end{tabular}
\caption{Preparation parameters and characterisation of the crystals.
The flux composition is given in the first block, followed by the
preparation atmosphere and the temperature program,
with a slow decrease at the indicated rate in the given
interval, and a quench or slow cooling to RT. The morphology and
possible growth twins are indicated in the last block.
}
\label{prepcond}
\end{table}

The two crystals prepared to obtain off-stoichiometric $x$ were
oxidized in 1 bar flowing oxygen in the temperature range
600--300$^\circ$C to obtain $y=1$. The crystals of the series with
various oxygen content were prepared along the route denoted $x=0$ in
the table, the oxygen content was set in a similar tempering step
under suitable oxygen pressure. The oxygen concentration $y$ expected
from the tempering conditions is in all cases in accord with the
relative NMR-line intensities of the Cu(1)-O-coordinations in the
chains (see below).  The error in $y$ estimated from the NMR-spectra
is 5\%. We expect that we may not reach the end members (especially
$y=0$) by approximately that margin, but will nevertheless denote them
for brevity with $y=0$ and $y=1$ respectively. We did not succeed to
prepare superconducting Pr123 along this route, all samples are
semiconducting.

At this point we would like to note that the local Pr/Ba-ratio of a
given sample with fixed overall $x$ can be influenced by an additional
heat treatment at higher temperature as well, but the composition of
the resulting crystal is inhomogeneous. For small deviations from the
equilibrium surface a crystal may decompose into a member of the solid
solution system with a different RE/Ba ratio, and either other
stable phases like BaCuO$_2$, or a second member of the solid solution
system. Larger deviations from equilibrium may lead to a spinodal
decomposition where, for example, the Pr/Ba-ratio varies laterally and
changes with time. The final state of these decomposition reactions
depends on the initial RE/Ba-ratio and whether or not this cation
ratio is located inside the four-dimensional volume of the solid
solution system. In the case of a spinodal decomposition the regions
of different Pr/Ba ratios within a sample are very close to each other
(10 - 100 nm), and any integral chemical analysis will reproduce the
composition of the untreated sample.

Unfortunately, the substitution of Ba by Pr (abbreviated $[Pr]_{Ba}$
below) is hard to detect by standard scattering techniques because of
the similar scattering cross sections of Pr and Ba. We have,
therefore, up to now in our preparation little quantitative control of
$x$. A unique way to determine the $[Pr]_{Ba}$-concentration is the
magnetic signal in polarized neutron scattering from these
defects.\cite{cha455} Crystals prepared along the route denoted $x=0$
were analyzed by Markvard et al. using this technique, and a
$[Pr]_{Ba}$-concentration near 5\% was detected.\cite{mar336} We
therefore expect $x\approx 0.05$ in the crystals we tempered to obtain
various oxygen contents.

Neutron diffraction showed that Pr on Ba-sites is not responsible for
the extraordinary suppression of superconductivity in Pr123. Kramer et
al. report that $[Pr]_{Ba}$ and $[Nd]_{Ba}$ have similar influence on
the materials properties.\cite{kra424} The RE-ion binds oxygen on
otherwise empty neighboring O(5)-oxygen sites between chains,
disrupting the chain order in the process.\cite{boo433} The defect is
expected to reduce the hole count in the planes and provides a strong
scattering center in the chain layer, where it may contribute to
suppress 1D-conductivity in the chains.\cite{yan160,fis457,yos456}
However, the experiments indicate that the influence of the defects is
limited to the chains, an independent mechanism not active in Nd123
must be present in Pr123 to suppress the MIT and superconductivity in
the CuO$_2$-planes.\cite{kra345} We emphasize, however, that
$[Pr]_{Ba}$ may well influence the low temperature magnetic structure
of the Cu(2)-sublattice by inducing magnetic moments on
Cu(1)-sites.\cite{uma400,and342} NMR and neutron diffraction
consistently showed that Cu(1)-moments induced by (nonmagnetic) Al
couple the planes ferromagnetically along the $c$-axis and can induce
a rotation transition of the Cu(2)-structure to avoid
Cu(1)-frustration.\cite{lon227,bre350} A similar magnetic transition
in Al-free, semiconducting Nd123 may well be due to Nd on
Ba-sites,\cite{dro304,boo00} and Rosov et al. reported ordered Cu(1)-moments
in Pr123 at low temperature.\cite{ros39} Magnetic Cu(1) is not
detected in Cu(1)-NQR (see below and ref.\cite{gre354}) and is,
therefore, most probably localized at defect sites in the structure.

While $x$ is always positive for Nd, the accessible range in the Pr/Ba
solid solution system may extend to the Ba-rich side, as is the case
for La/Ba in La123.\cite{lin463,gui232} Substitution of Pr (+3) by Ba (+2)
should increase the hole count in the planes, and in fact it has been
suggested that the superconductivity in Pr123 crystals prepared by Zou
et al.\cite{zou328,zou353,bla355} might be due to heavy Ba-doping of
the Pr-sublattice.\cite{zou353,nar426} Note that disorder on the
Pr-sublattice at smaller defect concentrations has been invoked as a
viable mechanism to localize the hole states in the
$4f-2p\pi$-hybridization band.\cite{lie197}

The symmetry and the morphology of the oxidized crystals depend in a
characteristic way on $x$ that we also find in La123 and
Nd123.  Ba-rich and stoichiometric crystals exhibit an orthorhombic
structure with twins, whereas the Pr-rich crystal kept its tetragonal
structure even after additional oxidation treatments under high oxygen
pressures. This result was expected because the extra oxygen ions on
the O(5)-sites between the Cu(1)O-chains reduce the orthorhombicity.
The morphology of the crystals changes with increasing RE/Ba-ratio
from isometric blocks, sometimes with additional (101)-faces, formed
by stoichiometric samples, to the formation of (100)/(001) growth
twins and, finally, even to (001)/(111) growth twins at large $x$.
All orthorhombic crystals investigated in this work are twinned with
respect to the $a$- and $b$-axes, but we find no indication for $c$-axis
twins in the NMR-spectra.

\begin{figure}
\psfig{file=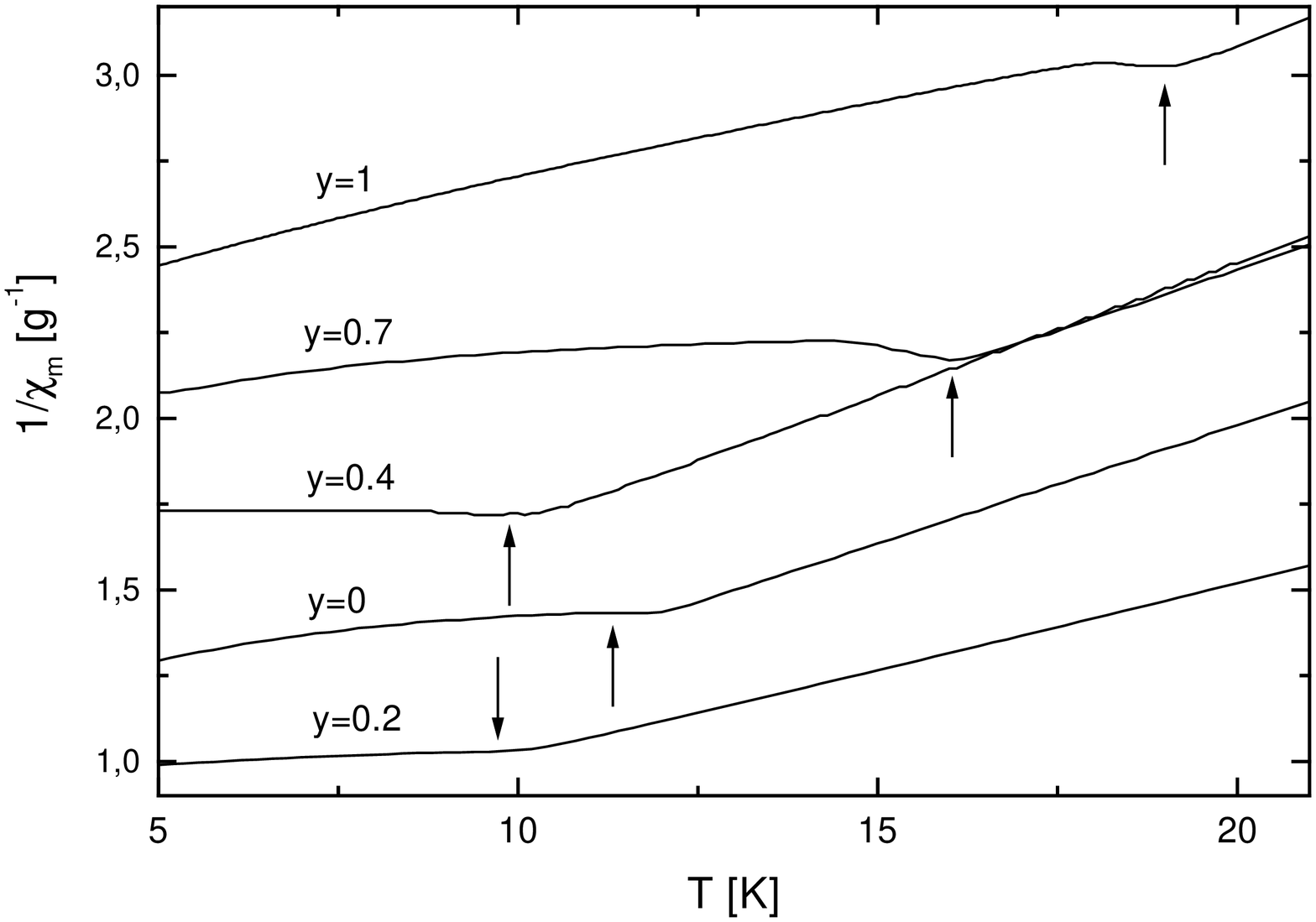,width=8cm,clip=}
\caption{
  Inverse low temperature $a,b-$plane mass susceptibilities showing
  the low temperature magnetic transition. The Cu(2)-sublattice orders
  antiferromagnetically near room temperature. Below $T_{N2}$ (arrows)
  Pr carries a moment of $0.3\ldots 0.9
  \mu_B$.\protect{\cite{moo408,li52,boo306}} The transition is very
  sensitive to sample quality, Uma et al. observed a transition in two
  steps in high-purity crystals grown in
  BaZrO$_3$-crucibles.\protect{\cite{uma419}} }
\label{Suszept}
\end{figure}

In fig.\ref{Suszept} the low temperature susceptibility in a field of
1~T perpendicular to the $c-$axis clearly displays the cusp associated
with the reorientation transition in the Cu(2)-sublattice and the
appearance of the Pr-moments at various oxygen concentrations.
The size and shape of the feature is very sensitive to sample
quality and even depends on magnetic history of the
sample.\cite{nar425} In clean samples the transition temperature
drops from near 17~K at $y=1$ to 12~K at $y=0$, similar to what we
find in our crystals. The transitions occure at 14.9~K and 12~K for
the Ba-rich and the Pr-rich sample respectively (not shown).
 
\subsection{The Cu(1)-O-Chains}

\begin{figure}
\psfig{file=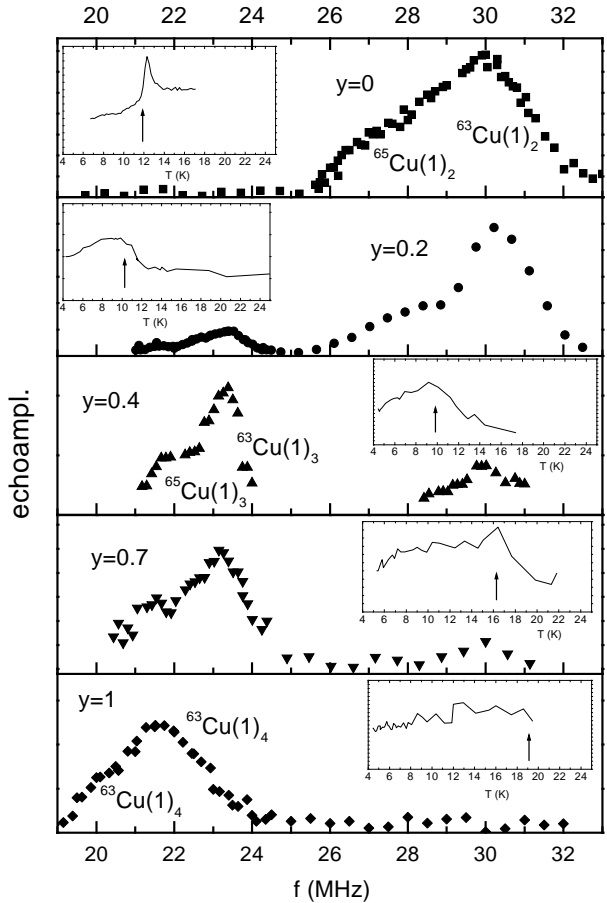,width=8cm,clip=}
\caption{Zero-field NQR spectra of the Cu(1)-sites in Pr123. The
  assignment of the lines to the two- three- and fourfold oxygen
  coordination of Cu(1) corresponds directly to the one for the other
  RE123.The insets show the temperature corrected echo amplitude at
  constant excitation conditions versus temperature to illustrate the
  systematic change in the influence of the fluctuations at $T_{N2}$
  (arrows, see text).}
\label{Cu1H0}
\end{figure}

\begin{figure}
\psfig{file=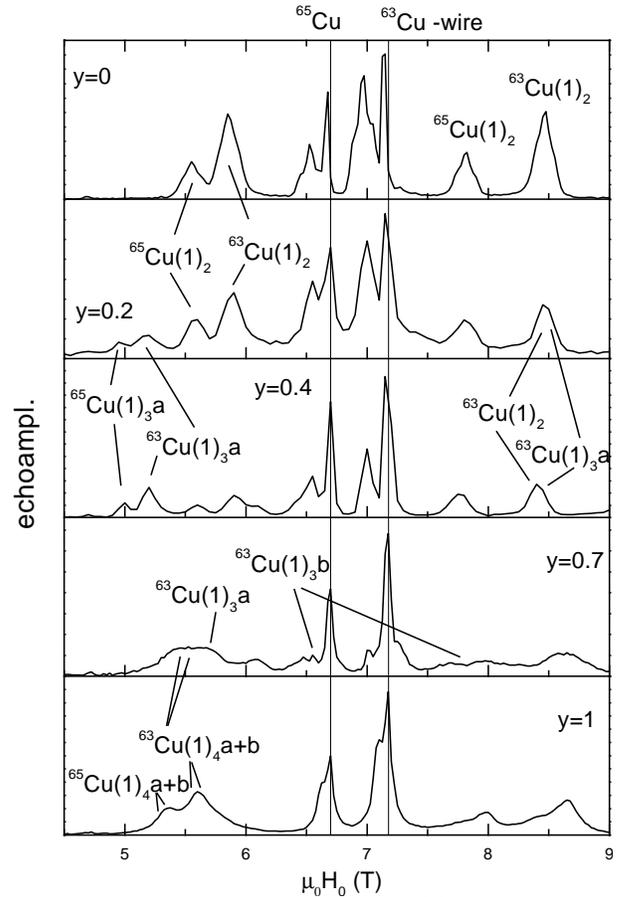,width=8cm,clip=}
\caption{Field-sweep spectra of the Cu(1)-sites in Pr123 at 81~MHz and 4.2~K 
  with the field along the (twinned) $a$- and $b$-axes. Together with
  similar spectra with the field along the $c$-axis (not shown) the
  positions of the quadrupole satellite transitions determine the
  electric field gradient tensor, their intensities the distribution
  of oxygen in the chains.}
\label{Cu1Hsw}
\end{figure}

Fig.\ref{Cu1H0} shows the zero field spectra, fig.\ref{Cu1Hsw} the
field sweep spectra in Pr123 at various oxygen concentrations $x$. The
spin-echo spectra were taken at 4.2~K and at comparably long pulse
distances ($\approx 40 \mu$s) to suppress the contribution of the
Cu(2)-sites.  The frequency range of 19 to 33 MHz for the NQR-spectra
of the Cu(1)-sites in the antiferromagnetic RE123 (RE$\ne$~Pr) and the
assignment of the lines to the different oxygen coordinations of
Cu(1) is by now well established.\cite{lue329,hei3} The line
positions and the relative intensities are similar to the ones
observed in the other RE123, so we can safely use the same assignsment
(Cu(1)$_4=21.8$~MHz, Cu(1)$_3=23.3$~MHz, and Cu(1)$_2=30.1$~MHz for
$^{63}$Cu). The subscripts denote the number of oxygen in the
Cu(1)-coordination shell, four for Cu(1) in a full chain, three for a
chain end position, and two for the apex oxygen in an empty chain
position.

The lines are significantly broader than in the other RE123 systems,
where widths well below 100 kHz are observed in the homogeneous end
members. The broadening may be due to structural defects inducing a
distribution in the electric field gradients (EFG), and to
inhomogeneities in the magnetic structure of the Cu(2)-lattice
inducing inhomogeneous internal magnetic fields. The two cases can be
distinguished from spectra obtained in a large external field, where
the frequency of the central transition is to first order independent
of the EFG, so its linewidth is dominated by inhomogeneities of
internal magnetic fields, while the distance between the satellites
and the central ransition is determined by the component of the
EFG-tensor along the magnetic field, so the satellites are broadened
by both, distributions in the EFG and in the internal magnetic field.
In fact, both cases have been observed, Nehrke et al. found a
significant narrowing of the Cu(1)-lines above $T_{N2}$ at $y=0$, a
clear indication that magnetic disorder is present at low
temperatures, while Grevin et al. confirmed our result that the
magnetic transition at $T_{N2}$ does not influence the linewidth at
$y=1$, instead they find that it is due to the formation of a charge
density wave at 100~K\cite{neh199,gre354}

The fluctuations at low temperature confirm, however, that the
transition at $T_{N2}$ in our crystals is magnetic.\cite{neh199} The
insets to fig.\ref{Cu1H0} display the systematic change of this behavior
with increasing oxygen content. The insets show the
$^{63}$Cu(1)-signal intensity, corrected for the Curie-law of nuclear
magnetization, versus temperature, measured across $T_{N2}$ (arrows).
The spin echo is measured at low temperatures with a repetition time
shorter than the relaxation time $T_1$ for equilibration of the
nuclear spin system. Under such circumstances one expects that the
amplitude increases when the relaxation time decreases, as is expected
near a magnetic phase transition. The stoichiometric crystal at $y=0$
(top) clearly shows this behavior. $T_1$ decreases with
increasing temperature and a pronounced minimum of $T_1$ at $T_{N2}$
leads to a peak in signal intensity. With oxygen doping $T_1$ becomes
shorter in the low temperature phase, leading to a higher signal than
above T$_{N2}$, and the relaxation induced peak at the transition is
lost, presumably swamped by the large background of fluctuating fields
in the ordered phase. At $y=1$ finally, no significant effects of the
transition could be detected in this simple way. However, we found
evidence for two contributions to the fluctuations in this temperature
range for the fully oxidized crystal in full $T_1$-measurements. One
component is almost independent of temperature and may be quadrupolar
or magnetic, the other shows a broad minimum at $T_{N2}$ and is
ascribed, therefore, to magnetic fluctuations.\cite{nehrkediss}

The point symmetry of the Cu(1)-sites implies that the principal axes
of the EFG-tensor for all oxygen coordinations are along the
crystallografic axes. The full tensor is, thereby, determined by the
assignment of the eigenvalues $V_x, V_y, V_z$ (in ascending order of
absolut values) to the axes, by the size of the largest eigenvalue
$V_z$, and by the asymmetry parameter $\eta=\abs{V_x-V_y}/V_z$. We
determined these values from fits to the field-sweep spectra with the
field along the $c-$axis, and along the $a-, b-$axes. The values are
compared in table \ref{Cu1EFG} 

\begin{table}
\caption{Comparison of the largest eigenvalue $V_{z}$(in $10^{-21}V/m^2$), 
the asymmetry
parameter $\eta $, and the eigenvalue of the EFG along the $c-$axis, for
the three Cu(1)-coordinations in Y123 and Pr123.}
\label{Cu1EFG}
\begin{tabular}{l|ccc|ccc}

& \multicolumn{3}{c}{RE=Y} & \multicolumn{3}{c}{RE=Pr} \\
Site       & $V_z$ & $\|c$  & $\eta$ &  $V_z$ & $\|c$  & $\eta$  \\
\tableline
Cu(1)$_2$  & 11.81 & V$_z$ & 0.0     & 11.74 & V$_z$ & $<0.05$ \\
Cu(1)$_3$  & 9.25  & V$_y$ & 0.3     & 8.84  & V$_y$ & 0.4     \\
Cu(1)$_4$  & 8.4   & V$_x$ & $>0.98$ & 7.9   & V$_x$ & $0.9(\pm .1)$ \\ 
\tableline
\end{tabular}
\end{table}

\noindent
with the ones observed in Y123. We
cannot distinguish between $a-$ and $b-$axis in our twinned crystals,
therefore we give only the EFG-component along $c$. The tetragonal
symmetry of Cu(1)$_2$ implies $\eta=0$, as is observed.

\begin{figure}
\psfig{file=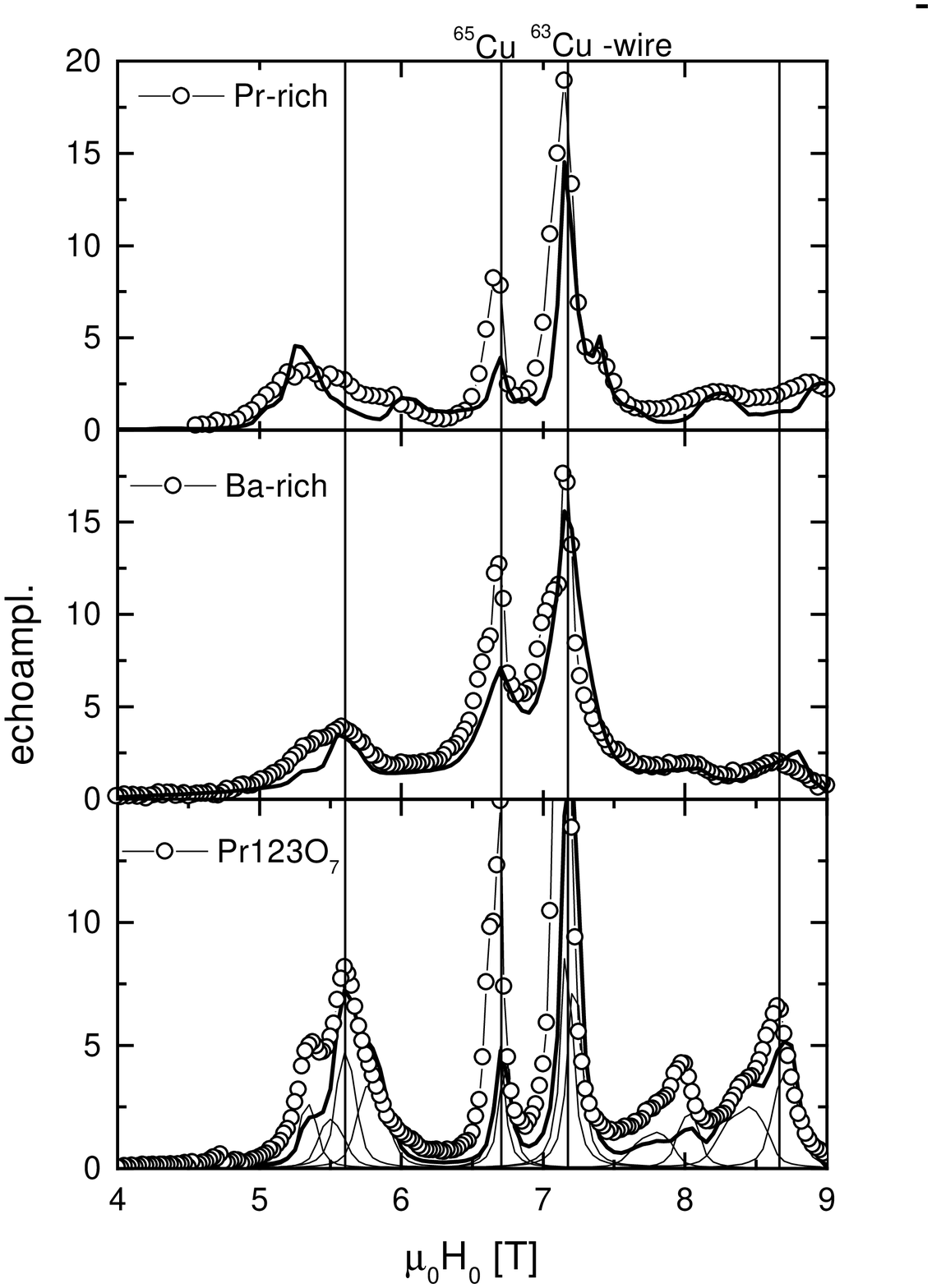,width=8cm,clip=}
\caption{Comparison of the field-sweep spectra of Cu(1) at 81 MHz in
  Pr123 at $y=1$ prepared in different oxygen atmospheres (see
  tab.\protect{\ref{prepcond}}). The vertical lines at the center
  transitions mark the Cu-resonance from the wire of the pick-up coil,
  and at the $^{63}$Cu-satellites the position in the stoichiometric
  sample (bottom). The full lines are simulated spectra (see text for
  parameters). For the stoichiometric sample the decomposition into
  four subspectra (two isotops in a field along the local $x$- or
  $y$-axis of the EFG) is also indicated.}
\label{DefPr}
\end{figure}

In fig.\ref{DefPr} we compare the field-sweep spin echo Cu(1)-spectra
of the three fully oxidized crystals, again with the external field
applied along the $a$-, and $b$-axes. The fourfold coordination
Cu(1)$_4$ characterized by the EFG-tensor
$\|V_{zz}\|\approx\|V_{yy}\|$ or $\eta\approx 0.9$ is the only one which we
could identify in our crystals, as is expected for filled chains.  All
Cu(1)-spectra are centered at the Cu-resonance of the metal, showing
that the Cu-chain sites detected in NMR carry no moment in these
crystals. We emphasize that this means only that there is no
homogeneous magnetization on the Cu(1)-sublattice. Cu(1)-moments
localized at defect sites in the structure may well be undectable by
NMR due to short relaxation times.

The spectrum of the Pr-rich sample (top) shows a larger splitting than
the stoichiometric one (bottom), an inhomogeneous broadening of the satellite
lines, but no significant broadening of the central line.  The high
field shoulder at 7.4~T of the $^{63}$Cu-central transition can be assigned to
the (111) growth twins mentioned above. As discussed above, the
broadening of only the satellite lines shows directly the presence of
disorder in the charge distribution of the chain layer, clear evidence
for the substitution of divalent Ba by tetravalent Pr and the
disorder this defect site induces in the oxygen chains. On the other
hand, the unchanged central line width proves the absence of internal
magnetic fields at the Cu(1)-site, indicating that the magnetic
structure of the Cu(2)-sublattice is not affected by the defects.

The spectrum of the nonstoichiometric Ba-rich crystal (center) is, in
contrast, best described by two subspectra. One with app. 30\%
relative intensity is the same as for the stoichiometric crystal,
accordingly the environment of 1/3 of the Cu(1)-sites is unchanged. A
significant broadening of the central transitions together with a
severe broadening of the satellites shows the presence of
inhomogeneous magnetic fields together with charge disorder for the
majority of the Cu(1)-sites. The inhomogeneous width of the central
transition corresponds to internal magnetic fields of $\approx 0.2$~T,
in full agreement with the transferred hyperfine fields from the
Cu(2)-moments at the Cu(1)-site ($\approx 0.1$~T) that are known from
studies of the so-called AF-II structure in Al-doped
Y123.\cite{bre350} The defects in this crystal do, therefore,
introduce inhomogeneities in the magnetic structure of the
Cu(2)-sublattice.

The most plausible explanation for the fact that the size of the
internal fields agrees well with the known transferred hyperfine field
from the Cu(2)-moments is that the divalent Ba dopes a hole state into
the $d_{x^2-y^2}-2p\sigma$-band of the CuO$_2$-planes, similar
to the well known case of Ca-doping on the RE-site in RE123.  One
might expect that this will form at low Ba-concentrations in the
RE-sublattice a localised Zhang-Rice singlet, which in turn would lead
to a Cu(2)-moment missing in the configuration of the neighbouring
Cu(1)-site. The intrinsic superconductivity of Pr123 crystals prepared
under special conditions discussed recently is in view of these
results most probably due to these holes becoming mobile at higher
Ba-concentrations, just as in the other high-$T_c$-cuprates.

\subsection{Pr-Resonance}

\begin{figure}
\psfig{file=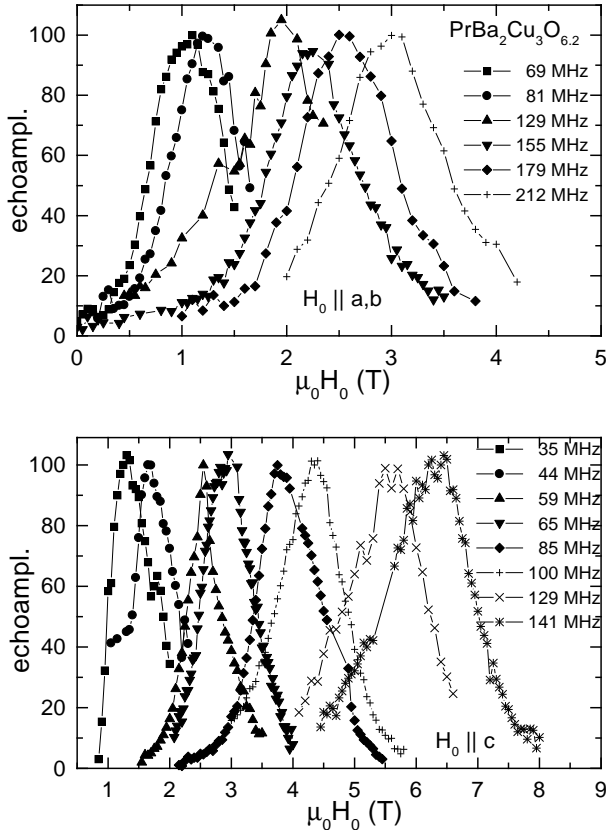,width=8cm,clip=}
\caption{Field-sweep spectra of Pr in slightly doped ($y=0.2$) Pr123
with the field along the c-axis (top), and the a- and b-axis (bottom).
At frequencies below $\approx 40$~MHz there is increasing overlap with
the Cu(1)-spectra, especially Cu(1)$_2$ at small oxygen concentrations.}
\label{Prx2cab}
\end{figure}

\begin{figure}
\psfig{file=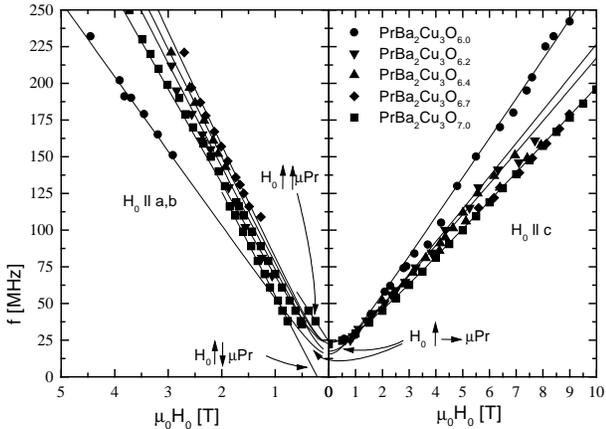,width=8cm,clip=}
\caption{Field dependence of the Pr-resonance frequency with the field
  applied along the c-axis (right), and along the a- and b-axes (left)
  for the different oxygen concentrations $y$. The full lines were
  calculated for the van Vleck paramagnetism of the Pr-$4f$-shell in
  the crystal field as described in the discussion. See Nehrke et al.
  for a detailed discussion of the line splitting in small in-plane
  fields.}
\label{PrfvH}
\end{figure}

We observed Pr-resonances very similar to the ones reported in
ref.ref.\onlinecite{neh352} in all crystals except one specimen with the nominal
concentration $y=0$. In a second reduced crystal we found the signal,
but its intensity is very small. Fig.\ref{Prx2cab} shows the field sweep
spectra at $y=0.2$, fig.\ref{PrfvH} shows the gyromagnetic lines for all
samples. The signal 

\noindent
is clearly due to $^{141}$Pr since this is the only
ion in the structure which may show van Vleck paramagnetism and the
corresponding anisotropic enhanced gyromagnetic ratios $\gamma_{eff}$
(the slopes at high fields, see below).

$\gamma_{eff}$ is almost indendent of $y$, only for $y=0$ we find a
higher value than in doped crystals with the field along the $c-$axis,
and a relatively small one with the field in plane. The fully
oxidized, stoichiometric crystal is the only one where an orthorhombic
distortion is detected by a splitting of the Pr-resonance line at low
external in plane fields (see ref.\onlinecite{neh352} for a detailed
discussion of this effect).  This does not prove, however, that the
symmetry is tetragonal on a local scale for $y\le 0.7$ in our
crystals. A splitting similar to the one reported for $y=1$ might be
hidden, because the smaller Pr-signal at lower $y$ overlaps with the
very complicated Cu(1)-spectrum. The linewidth is seen to be roughly
proportional to the frequency or the optimum field, indicating that
the inhomogeneous broadening is due to a distribution of the local
gyromagnetic ratios with a width of app. $\pm 10\%$. There is no
indication of any unresolved quadrupole splitting like the one
observed in Pr$_2$CuO$_4$,\cite{bak441} presumably because the EFG is
even smaller at the nearly cubic Pr-site in the Pr123 structure. We
point out that the Pr-linewidths turn out to be independent of oxygen
content, in contrast to the Cu(1)-resonances.

\begin{figure}
\psfig{file=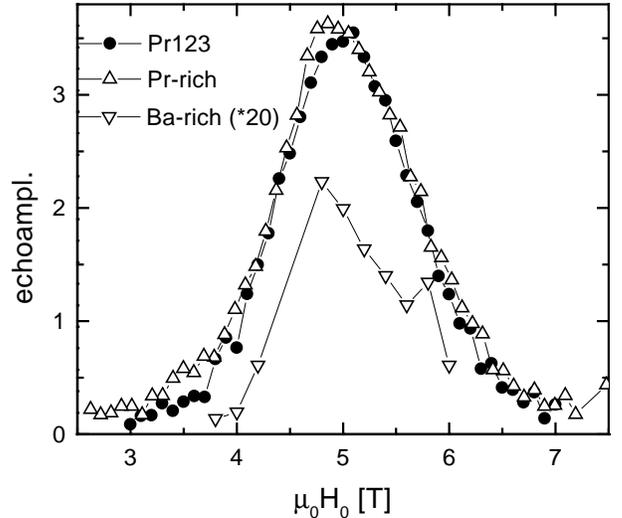, width=8cm, clip=}
\caption{Pr-field-sweep spectra at 102~MHz, 1.3~K, in a field $B_0\|c$. The 
amplitudes are scaled to allow comparison of the line shapes. The signal
in Pr$^-$123 is smaller by more than one order of magnitude.}
\label{PrRes}
\end{figure}
 
Comparison of the Pr-spectra in the three fully oxidized crystals in
fig.\ref{PrRes} shows immediately that the effective gyromagnetic
ratios and the linewidths of the stoichiometric and the Pr-rich sample
are identical. The intensities of the two signals have been scaled to
the same maximum for this comparison. The result shows most clearly
that the local electronic state of the Pr-ions contributing to the
signal is the same in the stoichiometric and the Pr-rich crystal. As
indicated above, there is no additional broadening in the Pr-spectrum
for off-stoichiometric oxygen concentrations, in contrast to the
obvious inhomogeneous broadening of the Cu(1)-resonance. This gives
strong support to our earlier assignment of this signal to Pr on
regular RE-sites and not to the $[Pr]_{Ba}$-defect, because the
regular site is shielded from the structural disorder in the chain
layer by the CuO$_2$-planes.

In contrast to the similarities of the Pr-resonance on the Pr-rich
side of the solid solution system we have scarcely been able to detect
any Pr-resonance in the Ba-rich sample. Again this supports our
assignment of the Pr-signal, since Ba substituting for Pr introduces
disorder in the RE-sublattice and it has to be expected that large
changes in the effective gyromagnetic ratios of the neighbouring
Pr-sites will effectively wipe out the Pr-signal from that region.

\section{Discussion}

\subsection{Hole doping}

The hole doping mechanism of the CuO$_2$-planes with increasing oxygen
content of the chains in the RE123-structure must be analyzed in two
separate steps. First, the distribution of oxygen on the chain sites
has to be determined, then the number of holes induced by the various
Cu(1)-O-configurations has to be found. Summation over the cluster
probabilities times the corresponding hole counts should then yield
the overall hole concentration for a given oxygen concentration
$y$.\cite{mcc434} Cu(1)-NQR may, in principle, contribute to both
questions. The resonance frequencies determine the quadrupole
splittings and reflect the charge distribution, the relative
intensities of the lines correspond to the probabilities of the
different oxygen coordinations of Cu(1), so they give information on
the oxygen distribution.

The EFG at Cu(1)-sites depends sensitively on the charge distribution
on the chains. It is hard to calculate the EFG in a correlated
electron system from first principles, but for the chain layer local
density calculations show convincing results.\cite{sin135} The main
contributions arise from the charge distribution within the first
Wigner-Seitz cell. This justifies the phenomenological ansatz to
separate the EFG-tensor into a contribution from the incompletely
filled Cu-$d_{x^2-y^2}$ orbital and one from the $p$-orbitals of the
neighbouring oxygens. Ohno et al. use for the Cu(1)-site a
$3d$-contribution of 117~MHz/hole and $n_{4p}\nu_Q^{4p}=-67$~MHz for
the contribution from $n_{4p}$ holes on the four oxygen
neighbors\cite{ohn339}, making the quadrupole splitting very sensitive
to the charge distribution in the chain layer. Nevertheless, from
table \ref{Cu1EFG} the EFG of Cu(1) in Y123 and Pr123 is very similar
for all oxygen coordinations. $V_z$ in Pr123 is systematically smaller
by $\approx 5\%$, which might well be due to the differences in the
lattice constants, and $\eta$ of Cu(1)$_4$ is somewhat reduced in
Pr123, which might be connected to the relatively large defect
concentrations. Furthermore, we found no evidence that

\begin{figure}
\psfig{file=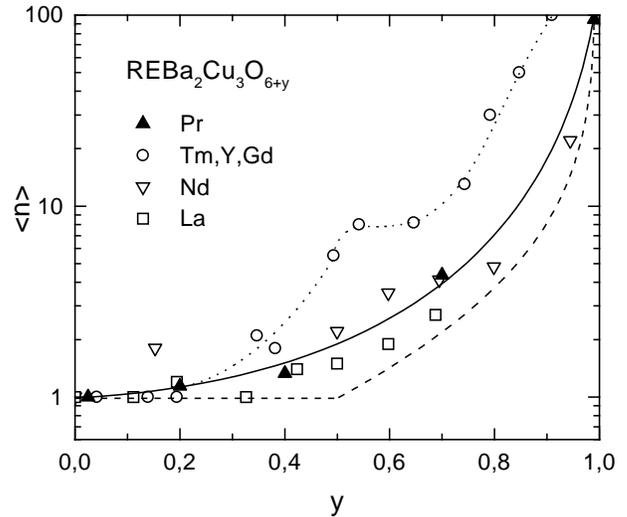,width=8cm,clip=}
\begin{center}
\psfig{file=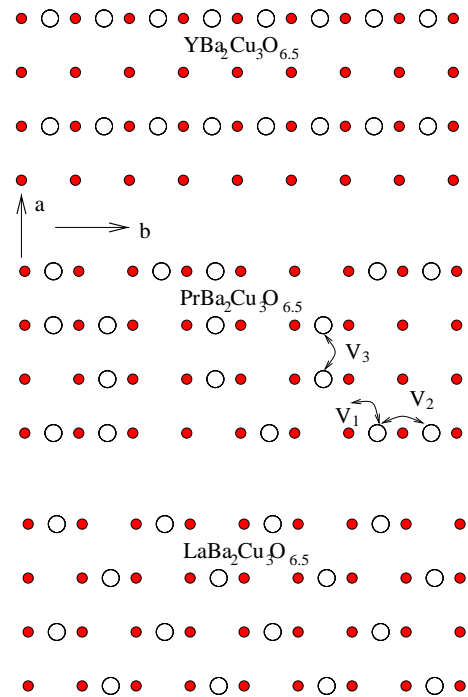,width=6cm,clip=}
\end{center}
\caption{Comparison of the mean Cu(1)-O-chain length versus $y$
  determined from the relative intensities of the Cu(1)-lines in RE123
  with RE=La, Tm, Gd, Nd, Y (\protect{\cite{lue329}}), and Pr (this
  work).  The full line sketches the expected behavior for a random
  distribution of chain oxygen, the dashed for repulsive O-O
  interaction $V_2$, and the dotted for attractive $V_2$. The bottom
  part shows the corresponding distributions of oxygen on the chains
  at half filling. This dependence of the oxygen distribution on the
  RE-radius may lead to a smaller doping of the CuO$_2$-layer at
  intermediate y for larger RE, but the influence of the RE-radius 
  on $T_{c max}=89\ldots 96$~K is only small.}
\label{Ostat}
\end{figure}
\noindent
the EFG at any
of the Cu(1)-coordinations depends on the oxygen content, in contrast
to the expectations in case of a localization of the carriers in the
1D-chains. We conclude that the results strongly support the ansatz
that the substitution of Y by Pr has very little influence on the
carrier concentration in the chain layer.

Cu(1)-lines from sites in CuO-clusters of different sizes are not
resolved, so a direct determination of the cluster probabilities from
the Cu(1) line intensities is not possible. Following the procedure
described by Heinmaa et al. we may estimate, however, the mean chain
length $\Erw{n}=1+2I_4/I_3$ from the relative intesities $I_4/I_3$ of
the three- and fourfold coordination.  Note that there is no model for
the oxygen distribution on the chain sites involved in this evaluation
of $\Erw{n}$. The small size of the crystals introduces a relatively
large error in the orientation which has a strong influence on the field
sweep spectra especially for Cu(1)$_3$. To check for consistency we
compare the intensities with the concentration $y=I_3/2+I_4$. If we
consider only spectra which could be fitted consistently in this sense
we arrive at the dependence of the chain length on $y$ compared in
fig.\ref{Ostat} with the corresponding results from L\"utgemeier et al.

Systematic errors for $\Erw{n}$ are due to the overlap of the broad
lines and to the assumption that no Cu(1)-contributions are missing in
the experimental spectrum. It is quite possible that part of the
Cu(1)-sites cannot be detected in NMR, either due to very large line
broadening from an inhomogeneous EFG-distribution, or due to fast
relaxation induced by slowly fluctuating on-site moments. As indicated
above an ordered moment has been assigned to the chain sites in Pr123
by neutron diffraction. We can exclude the presence of a small moment
homogeneously distributed over the whole Cu(1)-sublattice. At $y\ge
0.4$ we find no evidence for magnetic line broadening, and for $y=0$
the magnetic broadening of $\approx 0.2$~T together with the Cu(2)
hyperfine coupling constant in RE123 ($\approx 13.4 T/\mu_B$) sets an
upper limit of $0.02 \mu_B/$Cu for the crystals with low oxygen
concentrations. However, we cannot exclude the existance of localized
moments on Cu(1)-sites neighbouring defects. We emphasize again that
such defect induced moments determine most probably the low
temperature magnetism of the Cu(2)-sublattice in Al-doped
samples,\cite{schmenn} and they might well be present on
Cu(1)-neighbours of Pr on Ba sites.

Despite this uncertainty in the integrated intensity of the
Cu(1)-spectrum, which is common to some extent to all RE123-compounds
with light RE, our experimental chain lengths compare favourably with
the expectation for the chain length in the asymmetric next nearest
neighbour Ising (ASYNNNI) model with vanishing $V_2$, the interaction
of two oxygen nearest neighbours with Cu(1) in between (see bottom of
fig.\ref{Ostat}). The large, repulsive interaction $V_1$ drives the
formation of chains with diverging length at $y=1$ in the ortho-I
structure. The formation of the ortho-II superstructure of alternating
empty and long chains in Y123 at $y=0.5$ is due to an attractive $V_2$
in the case of a small RE. If $V_2$ is strongly repulsive the
so-called herringbone superstructure is stable at $y=0.5$, and
$\Erw{n}$ will be one up to that point. This case is apparently nearly
realized with the large La-ion on the RE-site. Nd with a radius in
between is well described by a random distribution of the oxygen on
chains, as is Pr. The origin of this influence of the RE-ion on the
interaction potential of oxygen in the ASYNNNI-model is not clear at
this time, but Pr123 fits smoothly into the trend. This finding
underlines the conclusion above that Pr has no extraordinary influence
on the electronic state of the Cu(1)O-clusters. 

The concentration of hole states in the CuO$_2$-Pr-CuO$_2$-trilayer
should then be at all oxygen concentrations similar to the one in
Y123, namely below the $\approx 0.35/$unit cell in (superconducting)
Y123 at $y=1$. Cluster and band structure calculations of the chemical
potential as a function of the chain length in Y123 indicate that only
clusters of a length above approximately 3 oxygen sites can induce hole states
in the CuO$_2$ planes.\cite{uim436} Therefore, the reduced average
chain length in Pr123 may lead to a somewhat smaller hole
concentration in Pr123 than in Y123 at the same $y$, at least if this
threshold behaviour occurs in Pr123 as well. Besides this small
difference we arrive at the conclusion that on average at most every
third Pr localizes a hole within the CuO$_2$-Pr-CuO$_2$-planes of the
structure.

\subsection{The Pr $4f$ ground state}

From the above arguments we are convinced that we can use the
Pr-resonance to probe the charge distribution in the
CuO$_2$-Pr-CuO$_2$-trilayer by its effect on the CEF of the
Pr-$4f$-shell. In cases where the NMR-signal of non-Kramers ions like
trivalent Tm or Pr can be observed, rather detailed information
especially on the low energy part of the CEF-Hamiltonian is
obtained\cite{bak441}. An experimental determination of the crystal
field at Pr from NMR is the more desirable since the spectra obtained in
Pr123 by INS are heavily broadened and the difficulties
in their analysis lead to a considerable scatter in the CEF-energy level
schemes\cite{hil147}. A more extensive description of our analysis of
the crystal field Hamiltonian $\Ham_{cf}$ than was possible in
ref.\onlinecite{neh352} seems to be in order in view of the controversal
discussion initiated by the results obtained for the crystal with the
highest oxygen concentration.

As stated above, the anisotropic enhancement of the gyromagnetic ratio
($^{141}\gamma=13$~MHz/T without van Vleck contributions)
unambiguously identifies the signal as due to $^{141}$Pr. It is,
however, more difficult to identify the position of the Pr in the
lattice, and it has been argued that our NMR-experiments might detect
the signal from the $[Pr]_{Ba}$-defect sites.

\begin{figure}
\begin{center}
\psfig{file=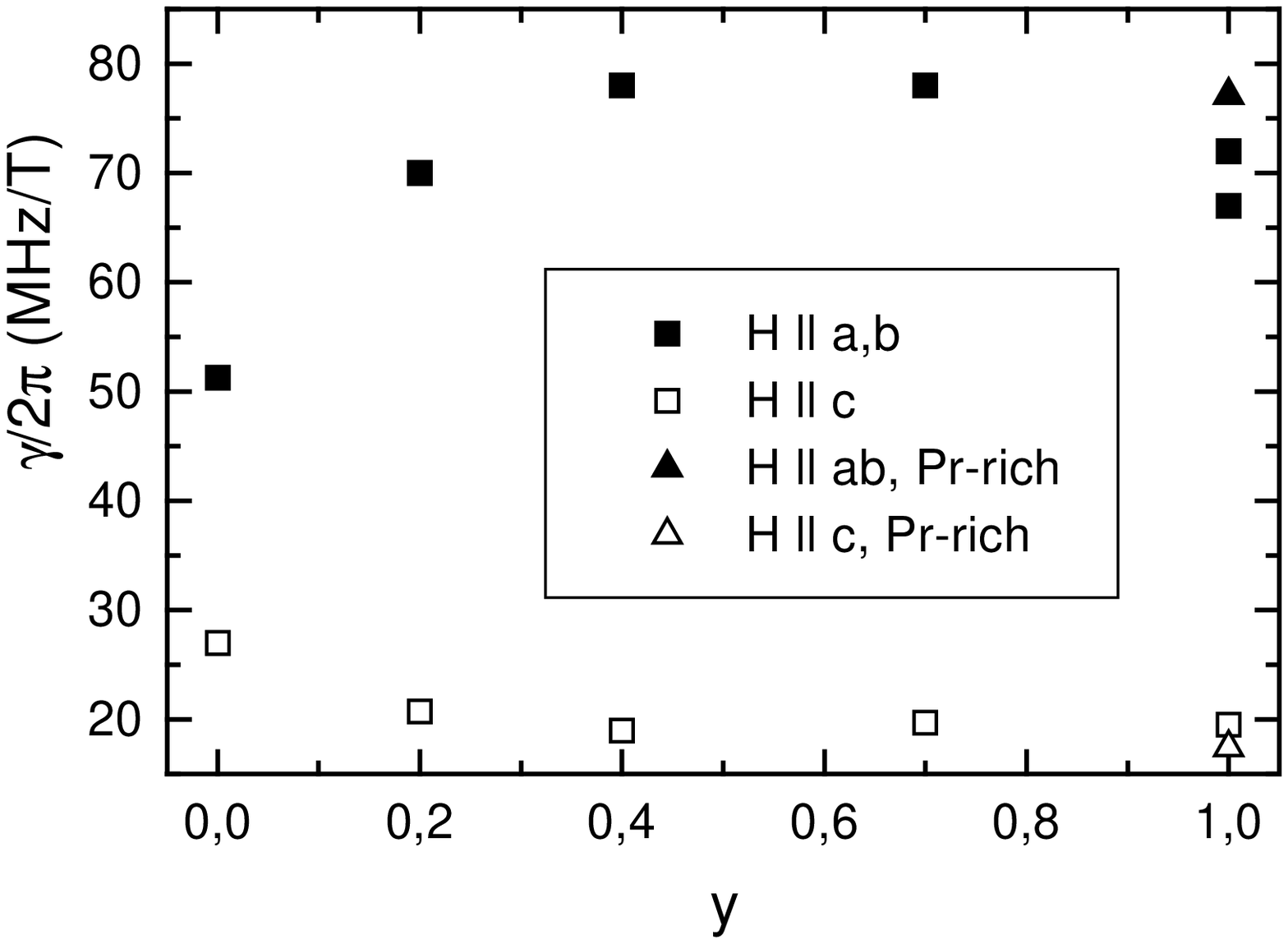,width=6cm,clip=} \\
\psfig{file=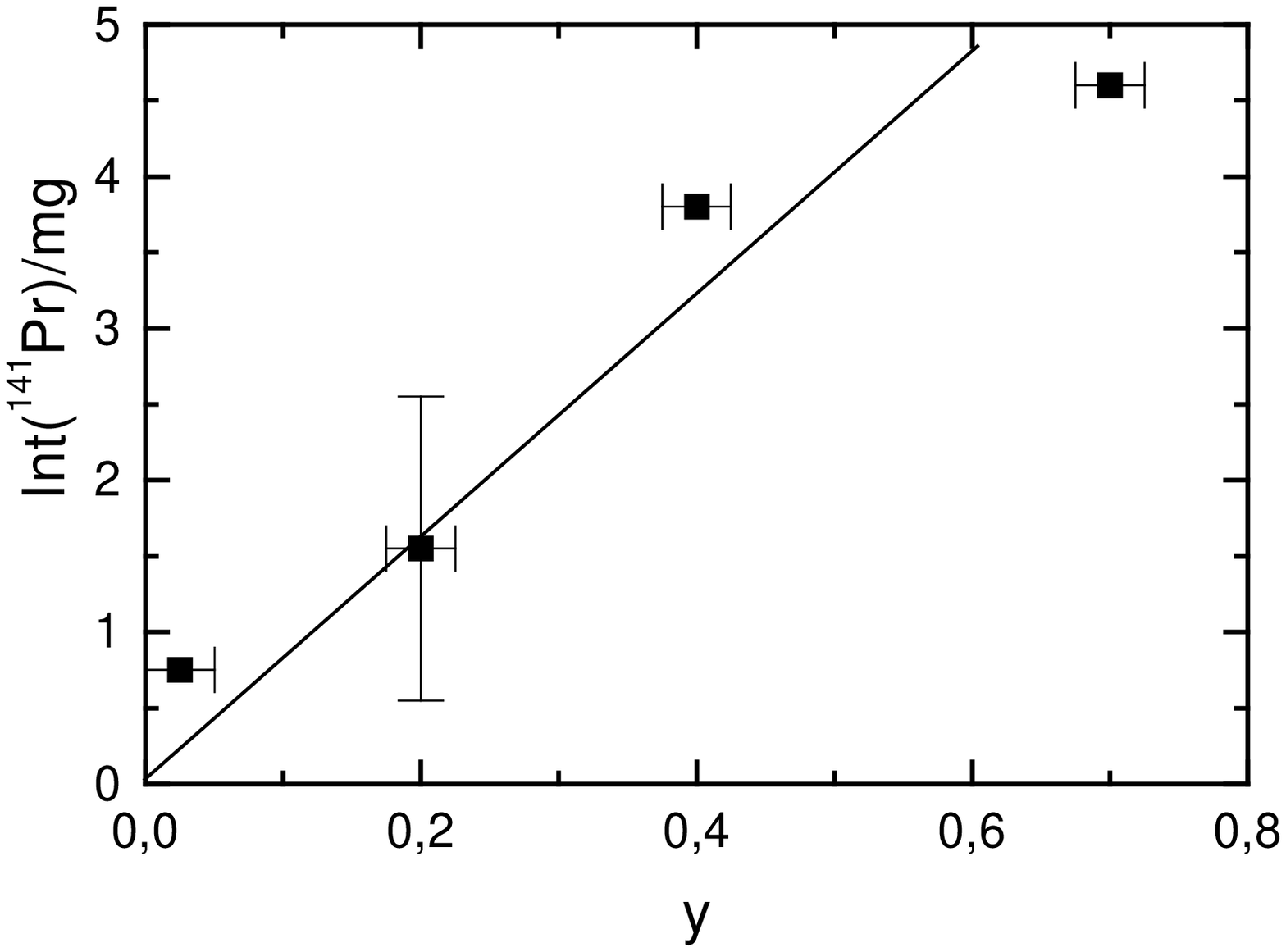,width=6cm,clip=}
\end{center}
\caption{Dependence of the effective gyromagnetic ratios of Pr (top),
  and the Pr signal amplitude normalized to sample volume (bottom) on
  the oxygen concentration. The error bars in the right figure
  indicate along $y$ the uncertainty from the relative Cu(1)-line
  intensities, along the intensity axis they estimate the uncertainty
  due to the different filling factors of the NMR-coils. The sample
  denoted throughout the text by $y=0$ for convenience is almost
  certainly not fully reduced. In a second crystal with the same
  Cu(1)-spectra we observed no Pr-resonance at all.}
\label{gamuIvx}
\end{figure}

There is strong experimental evidence against the assignment to the
defects. As indicated above, the linewidth of the Pr-resonance is
independent of any oxygen- or Ba-site disorder in the chains, though
this is clearly reflected in the Cu(1)-spectra at intermediate $y$ or
off-stoichiometric Pr/Ba cation ratio. Second, figs.\ref{PrfvH} and
\ref{gamuIvx} show in accord with the detailed crystal field analysis
below that the van Vleck susceptibility tensor is very nearly
tetragonal, and almost independent of $y$. This shows a tetragonal
symmetry of the crystal field (CEF), independent of the oxygen
contents of the chains. A symmetry close to tetragonal is in full
accord with the expectation for the regular Pr-site at all $y$,
because the site is far from the chain layer and shielded by the
CuO$_2$-planes. The CEF of Pr on Ba-sites, on the other hand, is
expected to reflect the orthorhombic symmetry at $y=1$, where the
neighbouring chains are filled, but also at small $y$, because
Pr$^{3+}$ is known to bind oxygen on the neighbouring
O(5)-sites. In addition, Allenspach et al. showed that the CEF
of $[Nd]_{Ba}$ is completely different from the one on the
RE-site,\cite{all349} while we show below that the CEF we find from
our results and the one determined from INS are rather similar. The
strong asymmetry of this configuration has been observed in the EFG of
La on the Ba-site.\cite{tro319}

The connection between the anisotropic $\gamma_{eff}$ observed in
fig.\ref{PrfvH} and the CEF can be understood in terms of a single-ion
description of the $4f$-shell. In ref.\onlinecite{neh352} we derive the nuclear
spin hamiltonian

\begin{eqnarray}
\nonumber
\Ham =
(\hbar\gamma+\frac{A_J}{g_J\mu_B}\chi_\alpha)I_\alpha\mu_0H_{0,\alpha}
+\frac{A_J}{g_J\mu_B}\Erw{M_\alpha}_{sp}I_\alpha \\
= \hbar\gamma_\alpha\mu_0H_{0,\alpha}I_\alpha
+\frac{A_J}{g_J\mu_B}\Erw{M_\alpha}_{sp}I_\alpha
\label{e.Hvveff} 
\end{eqnarray}

\noindent
for the nuclear spin $\I$ coupling with a gyromagnetic constant
$\gamma/2\pi=13.0$~MHz/T to the external field $H_0$, and via the
hyperfine coupling constant $A_J/h=1093$~MHz to the moment $\Erw{\M}$
in a 4f-shell with angular momentum $J$ and Lande-factor $g_J=0.8$.
$\chi_\alpha=\partial M_\alpha/\partial H_\alpha$ is the static van
Vleck susceptibility with the field along the crystallografic axis
$\alpha=a,b,c$, $\Erw{M_\alpha}_{sp}$ is the spontaneous Pr-moment
along one axis. The diagonalization of $\Ham$ for $H_0\|\M_{sp}$ or in
the case of large external fields is trivial in the notation with the
effective gyromagnetic ratio $\gamma_\alpha$ in the second part of
eq.\ref{e.Hvveff}. The gyromagnetic ratio $\gamma$ changes to 

\begin{equation}
\gamma_\alpha = \gamma + \frac{A_J}{\hbar g_J\mu_B}\chi_\alpha,
\end{equation}

which is independent of temperature and field as long as this is the
case for the van Vleck susceptibility. This holds for temperatures and
fields $k_BT, \mu_0H_0\Erw{\M}_\alpha\ll\Delta_\alpha$, where
$\Delta_\alpha$ is the smallest CEF-splitting of $4f$-levels connected
by $J_\alpha$. From the linear field dependence of the resonance
frequency in fig.\ref{PrfvH} it is clear that $\gamma_\alpha$ at 1.3 K
is indeed independent of the field up to at least 9 T (5 T for the
field in the plane). The offset and the slope of the gyromagnetic
lines determine immediately the spontaneous Pr-moment $\Erw{\M}_{sp}$
and the local susceptibility $\chi_\alpha$ respectively.

The frequency offset is clearly smaller than 25 MHz for all samples.
It corresponds to a very small ordered moment of below $h\nu/A_J=0.023
\mu_B/$Pr in all cases, more than one order of magnitude smaller than
the $0.3\ldots 1.2 \mu_B$ detected with neutron diffraction or
M\"ossbauer spectroscopy. It was perhaps this property which gave rise
to the most severe doubts about the interpretation of our spectra of
the fully oxidized sample.
 
In order to understand the origin of this discrepancy it is usefull
to discuss first the CEF. We compare the three experimental components
of $\chi_\alpha$

\begin{equation}
\label{e.chiexp}
\chi_{\alpha} = \frac{\hbar g_J\mu_B}{A_J}(\gamma_\alpha-\gamma)
\end{equation}

with the ones calculated in the eigenbasis of a suitably modelled
crystal field Hamiltonian $\Ham_{cf}$
($\Ham_{cf}\Ket{k}=$\\ $\eps_k\Ket{k}$):

\begin{eqnarray}
\label{e.chicf}
\nonumber
\chi_{\alpha}=  \\
\nonumber
(g_J\mu_B)^2\sum_k[\frac{1}{k_BT}\sum_l^{\eps_l=\eps_k}
|\Bra{k}J_\alpha\Ket{l}|^2\Exp{-\eps_k/k_BT}+ \\
\sum_l^{\eps_l\ne\eps_k}|\Bra{k}J_\alpha\Ket{l}|^2
\frac{\Exp{-\eps_k/k_BT}-\Exp{-\eps_l/k_BT}}{\eps_l-\eps_k}].
\end{eqnarray}

Because of the exponential factors in eq.\ref{e.chicf} the
susceptibility is dominated by the matrix element of the total angular
momentum with the ground state and the lowest excited state with
nonvanishing matrix element. The most convenient representation of
$\Ham_{cf}$ for a calculation of $\chi_\alpha$ is by means of the 
Stevens equivalent operators $\Ope_l^m$ with crystal field parameters
$b_{lm}$:

\begin{equation}
\label{e.Hcf}
\Ham_{cf} = \sum_{lm}b_{lm}\Ope_l^m.
\end{equation}

It describes the two $4f$-electrons in Russel-Saunders coupling. The
$LS$-admixture from spin-orbit coupling,\cite{Bleaney} which is
3100~K,\cite{McCausland} and the interconfiguration coupling, i.e. the
admixture of states from the multiplets $J=5, 6$ considered by
Hilscher et al.,\cite{hil147} are neglected. This may be justified as a first
approximation since the spin-orbit coupling is significantly larger
than the crystal field splitting. Then $\J = {\bf L}+{\bf S}$ is a
good quantum number and the ground state of an isolated Pr$^{3+}$ is
the ninefold $^3H_4$-multiplet, which is further split by $\Ham_{cf}$,
or by a magnetic field. The representation is based on the
Wigner-Eckhard theorem, which guarantees that the matrix elements of
$\Ham_{cf}$, taken within a subspace of states $\Ket{LSJM_J}$ with
fixed quantum numbers $L, S$, and $J$, are proportional to a suitable
polynomial $\Ope_l^m$ of angular momentum operators. The equivalent
operators and the numerical factors occurring in the Wigner-Eckard
theorem are tabulated by Hutchings\cite{Hutchings}.

The advantage of the equivalent operators $\Ope_l^m$ is that the
matrix elements can be calculated in a physically transparent way,
since they contain only angular momentum operators, and usually a good
fit of the energy spectrum is possible. We follow, therefore, the
common practice in NMR and use below equivalent operators. Note that
$A_J$ and $g_J$ enter both experimental parameters, the slope and the
offset of the gyromagnetic lines. The values for $A_J$ and $g_J$
introduced above are the ones for the $^3H_4$-multiplet, and they
allow for a fit to the gyromagnetic lines with reasonable parameters.
This is not possible if $A_J$ is chosen significantly higher (to allow
for a higher $\M_{sp}$), because then unphysically large moments result
in high field.

A good fit of the CEF energy spectrum from within the $J=4$-multiplet
alone does, however, not necessarily imply the same precision for the
crystal field parameters. The analysis of the INS-spectra is usually
based on a irreducible representation of $\Ham_{cf}$ by
spherical harmonics and shows admixture of states from higher
configurations. The conversion factors for the crystal field parameters
$B_{lm}$ used in the irreducible representations and the $b_{lm}$ above
may be determined by comparison of the coefficients with equal symmetry
and have been tabulated by Kassman\cite{kas220}. A comparison of
$\Ham_{cf}$ in the two representations is difficult if the
interconfiguration mixing has significant influence. In this case
differences may occure in the eigenvalues of $\Ham_{cf}$ calculated from
the irreducible representation and the ones calculated from the
equivalent operator representation after conversion of the parameters.

For $4f$-electrons spherical harmonics up to the order $l=2\cdot 3=6$ and
$m\le l$ can contribute to the series in eq.\ref{e.Hcf}. The point
symmetry of Pr allows only for even $l (=0,2,4,6)$ and $m (=0,2,4,6\le l)$
in orthorhombic point symmetry, and the fourfold $c-$axis in
tetragonal symmetry reduces this further to $l=2,4,6$ and
$m=0,4 \le l$. Obviously it is impossible to determine the full set from
the three components of the local susceptibility $\chi_\alpha$, and in
Pr123 even the more detailed information from INS is insufficient.
Fortunately, both methods are complementary: NMR is sensitive to the
ground state symmetry and the splitting of only the low lying levels,
while the INS-spectra suffer especially at low energy from the severe
broadening, but well resolved lines are observed $\approx 500$~K above the
ground state.

A simultaneous fit of $\Ham_{cf}$ to the results of both methods is
viable only for the samples with high oxygen concentrations $y$. The
difficulty is that our ground state symmetry for all $y$ must be
tetragonal with at most a very small orthorhombic distortion. The
severe line broadening of the INS-spectra does not allow to determine
the ground state symmetry unambiguously for $y=1$, but for $y=0$ any
description with tetragonal symmetry is very poor. We have, therefore,
apparently conflicting results from the two methods at $y=0$.

In order to solve the problem we propose to take the results of both
experiments at face value and assume the presence of Pr in two
different symmetries on the RE-sites. We emphasize that the presence
of two distinguishable electronic configurations of Pr was already
proposed by Fehrenbacher and Rice, so we use below their nomenclature
for the two sites. Here we only make use of the fact that the CEF
should be different for Pr$^{3+}$, the site without a hole state in
its $4f$-shell and in the $2p\pi$-orbitals of the eight oxygen ligands,
and for Pr$^{IV}$, the site where a hole is localized in the
$4f$-shell and the $2p\pi$-ligand orbitals. The concentration of the
Pr$^{IV}$-configuration is according to Fehrenbacher and Rice directly
the hole concentration $n_h$ determined above from the
Cu(1)-resonance, that is at most $\approx 0.35$ holes per unit cell or
roughly every third Pr-site.

First we assign the Pr-resonance detected in NMR to the
Fehrenbacher-Rice state Pr$^{IV}$, and Pr$^{3+}$ to the signal
obtained in INS (and most other techniques). From the NMR point of
view this is natural because the intensity of Pr-resonance increases
with $y$, that is, with the number of holes (see fig.\ref{gamuIvx}).
From the INS-results the assignment is necessary because a signal is
present at $y=0$ and the CEF observed is at all $y$ similar to the one
determined with very high accuracy for other RE in RE123, where the
hybridization state RE$^{IV}$ does not exist.

The contradicting results of INS and NMR for the reduced samples ($y=0$)
are now easy to understand. NMR observes a small signal from a few percent
of Pr in the state Pr$^{IV}$, which may be induced e.g. by a small
amount of oxygen remaining on chain sites. In INS the small signal
corresponding to these sites is probably not detected. On the other
hand, the Pr-resonance from the state Pr$^{3+}$ dominating the
INS-spectra may well be not observable in NMR, either due to fast
relaxation, or due to a large inhomogeneous broadening from the EFG of
the distorted $4f$-shell and the magnetic hyperfine fields induced by
$4f$-moments.

Near $y=1$, where app. $1/3$ of Pr is expected to be in the state
Pr$^{IV}$, they certainly contribute to the INS-spectra. Then we must
seek for a simultaneous fit of $\Ham_{cf}$ to the NMR-data and the
INS-spectra, weighted for their high sensitivity at low and high
energies in the spectrum of $\Ham_{cf}$ respectively. Any differences
in the CEF-energies of the two Pr-configurations must be compatible
with the supposedly inhomogeneous linewidth of the INS-spectra at
energies above app. 500~K.

The method we use to fit of $\Ham_{cf}$ to the NMR-data and the high
energy part of the INS-spectrum is motivated by the almost cubic
oxygen coordination of the RE-site in RE123. We start from a CEF with
cubic symmetry. We then improve this solution with a tetragonal
distortion, which we may 'polish up' with an additional orthorhombic
distortion and a local magnetic field.

\begin{figure}
\psfig{file=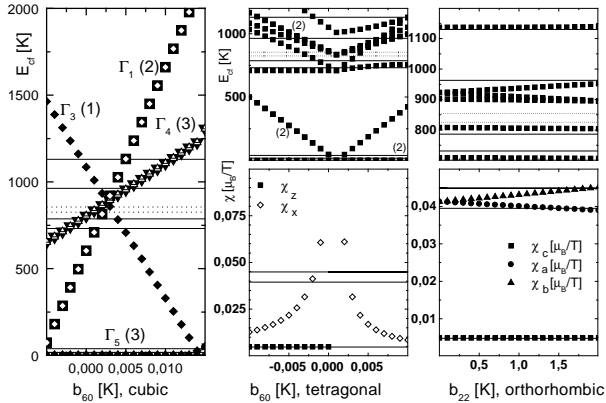,width=8cm,clip=}
\caption{Determination of the crystal field parameters $b_{lm}$ by
  successively lowering the symmetry in the fit from cubic (left), to
  tetragonal (center), to orthorhombic (right). The horizontal lines
  represent the crystal field splittings determined by INS (left and
  top figures) and the van Vleck moments of Pr determined from our
  NMR-data (bottom figures). Note that the upper right figure contains
  only the levels at high energies, showing that a reasonable fit to
  the energies in the resolved part of the INS-spectra is achieved
  with this simple procedure.}
\label{PrHcf}
\end{figure}

The three steps are shown in fig.\ref{PrHcf}. There is a general
consensus from all experimental techniques that the cubic
$\Gamma^5$-triplet is the subspace underlying the quasitriplet at low
energies of $\Ham_{cf}$ in Pr123. We therefore adjust the two
parameters $b_{40}, b_{60}$ of the cubic $\Ham_{cf}$ (see
ref.\onlinecite{lea000}) to fit the splitting between the ground state
and the other levels, and the one between the other levels roughly to
the experimental energies from INS. The left panel shows the regime of
$b_{60}$ for cubic symmetry, where the $\Gamma_5$-triplet is the
ground state ($b_{40}$ is fixed at 0.335 K).  We choose $b_{60}=0.001$
K as a cubic approximation.

Next we consider a tetragonal distortion by variation of $b_{60}$ with
$b_{64}=-0.021$~K fixed at its cubic value (center panels). The
$\Gamma^5$-triplet splits into a doublet and a singlet, which are the
ground state for $b_{60}$ larger or smaller than the cubic value
respectively (top part). The local susceptibility (bottom part)
$\chi_z$ fits the local susceptibility observed in the NMR-experiment
(horizontal lines) only if the singlet is the ground state. $\chi_z$
is too large by two orders of magnitude if $b_{60}$ is larger than its
cubic value (in the right half of the central panels). With the
singlet ground state the tetragonal $\chi_x$ ($\diamond$) falls
between the orthorhombic in plane susceptibilities at $b_{60}=-0.002$
K. Simultaneously the splitting of the upper levels is described more
or less correctly (top). At small $y$, where we find tetragonal
symmetry, this procedure leads to the parameters given in
tab.\ref{t.CFpar}.

The crystal $y=1$ is the only one where we found an orthorhombic
splitting, and here we introduce the orthorhombic distortion $b_{22}$
shown in the right panel of fig.\ref{PrHcf}. The changes in the high
energy eigenvalues are small (top, note the suppressed zero), and even
the eigenvectors of $\Ham_{cf}$ change only by a few percent. The fit
to the transitions observed in INS (horizontal lines) is already
reasonable. The local susceptibilities (bottom) are well described at
$b_{22}=1.9$~K. We may now use the other parameters to 'polish the fit
up', but we emphasize that this is far from unambiguous and any
improvement is probably beyond the accuracy of the description with
Stevens equivalent operators. The parameters given for $y=1$ in
tab.\ref{t.CFpar} are an example for an almost exact fit to the local
susceptibility {\em and} the energies above 500~K.

\begin{table}
\caption{Top: Crystal field parameters determined from the simultaneous
fit to the high energy INS-spectra and to the local susceptibilities
measured by NMR in the case of Pr123, $x=1$. Since we find tetragonal
symmetry for the local susceptibility we fit only to $\chi_\alpha$ for
Pr123, $y=0$, and searched for the nearest solution in the parameter
space to the one of $y=1$. Bottom: Eigenvalues of $\Ham_{cf}$ as
determined by Hilscher et al. from INS \protect{\cite{hil147}}, and from
our NMR-data. Values in brackets are not directly observed. Note that we
did not intend to fit the energies in the case $y=0$.}
\label{t.CFpar}
\centering{
\begin{tabular}{|c|ccccccccc|}
$b_{lm}$ [K] & $b_{20}$ & $b_{22}$ & $b_{40}$ & $b_{42}$ & $b_{44}$
& $b_{60}$ &  $b_{62}$ & $b_{64}$ & $b_{66}$ \\
\hline
$y=1$ & -1.9 & 1.2 & 0.325 & 0 & 1.665 & -0.0022 & 0 & -0.0252 & 0 \\
$y=0$ & 0 & 0 & 0.335 & 0 & 1.675 & -0.0069 & 0 & 0.05 & 0 \\
\end{tabular}
}

\centering{
\begin{tabular}{|l|ccccccccc|}
$E_{cf}(x=1)$     &   &    &    &     &     &       &       &     & \\ 
INS [K]           & 0 & 17 & 38 & 731 & 786 & (825) & (854) & 963 & 1130 \\
NMR [K]           & 0 & 132 & 142 & 726 & (775) & (832) & (922) & (956) & (1126) \\
\hline
$E_{cf}(x=0)$      &   &    &    &     &     &       &       &     & \\ 
INS [K] & 0 & 20 & 39 & 714 & (727) & (733) & 757 & 882 & 983 \\
NMR [K] & 0 & 208 & 208 & 351 & (506) & (835) & (866) & (866) & (1452) \\
\end{tabular}
}
\end{table}

\begin{figure}
\psfig{figure=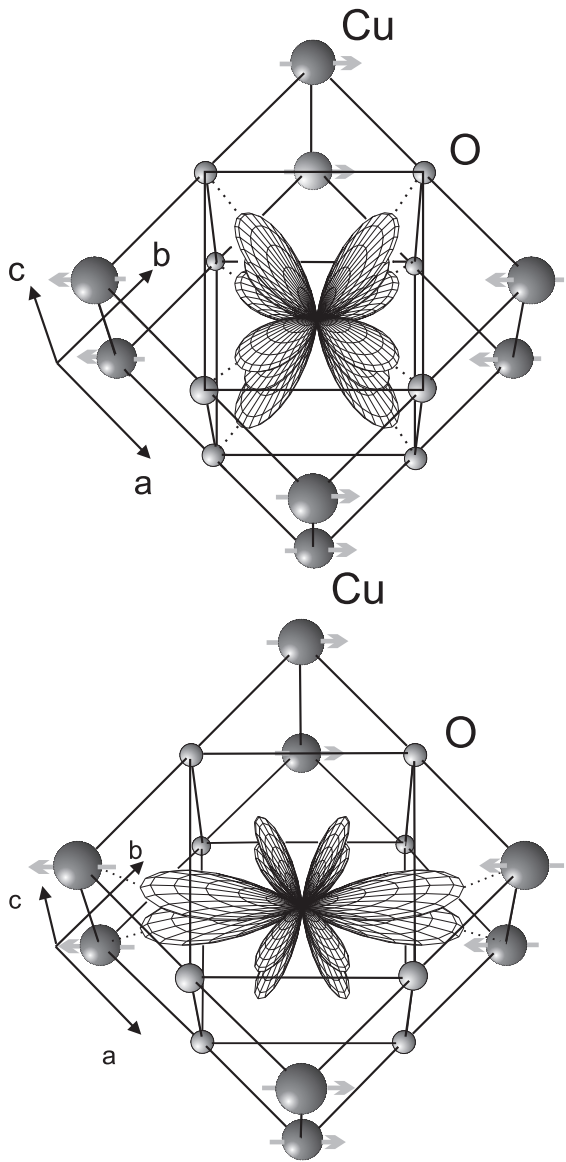,width=8cm,clip=}
\caption{Polar plot of the shape of the $4f^2$-wavefunction of Pr as
  determined from our NMR-data (top), and from INS (bottom). The
  (tetragonal) $c$-axis points to the top, the eight lobes along the
  local (111)-directions point to the oxygen neighbours. Only one of
  the two possible orthorhombic ground states for INS is shown, the
  other one is rotated by $\pi/2$ around the $c$-axis.}
\label{Pr4f}
\end{figure}

Fig.\ref{Pr4f} shows a polar diagram of the charge distribution in the
two-electron ground states calculated from our crystal field
parameters for $y=1$, and from the converted crystal field parameters
of Hilscher et al.\cite{hil147} The state Pr$^{IV}$ observed in NMR is
tetragonal, the small orthorhombic distortion is not visible.  The
eight lobes pointing along the space diagonals of (nearly) a cube with
oxygen at the corners resemble closely the $4f$-state envisaged by
Fehrenbacher and Rice\cite{feh198}. Note that there are four
additional lobes in the plane perpendicular to the $c$-axis pointing
to the edges of the cube. The ground state Pr$^{3+}$ observed in INS
looks very different at first sight, but from the above discussion the
connection between both states is easy to visualize: Rotations of
Pr$^{IV}$ by 90 deg around the local $x$- or $y$-axis ($\perp c$) lead
to the two remaining states of the cubic $\Gamma^5$-triplet, the
doublet in the tetragonal symmetry. In a suitable orientation of the
local $x$- and $y$-axes such a rotated Pr$^{IV}$-state resembles
closely the state Pr$^{3+}$ in the figure, only the four lobes which
were formerly perpendicular to $c$ are more pronounced.  Note that the
orientation of the local $x$- and $y$-axes or these four dominant
lobes is determined only when e.g.  a Jahn-Teller distortion lifts the
twofold degeneracy of this ground state in tetragonal symmetry.

The physical origin for the different ground state symmetries is in
our model the one missing electron charge in the $2p\pi$-ligand
orbitals of the oxygen coordination of Pr in the state Pr$^{IV}$. In
the presence of this hole the tetragonal CEF is on the left hand side
in fig.\ref{PrHcf} with the singlet ground state. If the orbitals are
occupied the CEF changes to the right hand side and the doublet
becomes the ground state.  The doublet degeneracy may be lifted by a
Jahn-Teller distortion leading to the ground state observed by INS.

\subsection{The low temperature magnetic structure}

We now turn to the low temperature magnetic state of Pr. All crystals
investigated here show the well known kink in the low temperature
susceptibility, indicating a magnetic transition at temperatures
increasing with oxygen content from $\approx 10$K to 18K (see
fig.\ref{Suszept}). In the discussion following our first observation
of nonmagnetic Pr it became clear that there is a simultaneous
transition in the magnetic structure of the Cu(2)-sublattice, but even
taking this complication into account moments of at least $0.5
\mu_B/$Pr are required to describe the magnetic Bragg peaks at all
$y$.

The observation of a Pr-moment is consistent with the interpretation
presented above if the state Pr$^{3+}$ carries this moment, and the
existence of two magnetically different sites has been considered much
earlier as a possible explanation for the broad $^{141}$Pr-M\"ossbauer
spectra.\cite{moo408} The fact that $T_N$ {\em increases} with the
concentration of the {\em nonmagnetic} site Pr$^{IV}$ is, however, in
clear contradiction to the concept that Pr-Pr exchange is the origin
of the transition. Even the fact that the transition occures at all
would be surprising at $y=1$, because the magnetism of the
RE-sublattice is known to be strongly 2-dimensional, and the
percolation threshold for the square lattice with nearest neighbor
exchange is only 0.5928. The 30\% nonmagnetic impurities introduced by
hole doping would bring the lattice almost to the percolation.
Obviously a similar difficulty with the exchange coupling arises with
the small dependence of the Pr-$T_N$ on magnetic dilution by other
RE-ions.

We propose, therefore, that the Pr-moments are induced by the
transition in the Cu(2)-sublattice. The mechanism is the same we
introduced in our previous work,\cite{neh352} but now we have to consider
two different Pr-sites. The different magnetic properties of the two
Pr-sites are a direct consequence of the two ground state symmetries
discussed above: The tetragonal symmetry of the state Pr$^{IV}$ has as
a consequence the cancellation of all transferred hyperfine fields
from Cu(2) at the Pr-site, therefore the Pr-moment vanishes. The small
moment indicated by the offset in the resonance frequency may be
induced by dipolar fields from the Cu(2)-moments, which do not fully
cancel by symmetry. 

On the other hand, the four pronounced lobes of the state Pr$^{3+}$
can point to the centre of the edges of the cube spanned by the
oxygen, that is to four of the Cu(2)-sites (see fig.\ref{Pr4f}). Note
that these moments are parallel in the structure proposed as a first
approximation in ref.\onlinecite{neh352} and sketched in the figure, and
that they do have a parallel component in the refined structure
determined later by neutron diffraction. In such a case the
transferred hyperfine fields at the Pr-site do not cancel. The size of
the induced moments depends exponentially on the small splitting of
the quasitriplet in the CEF (see eq.\ref{e.chicf}). Despite the
considerable scatter in the CEF-parameters determined for Pr$^{3+}$
from INS it is clear that the splitting is smaller by at least a
factor of two than the 130 K we obtained above for the state
Pr$^{IV}$, with the result that a transferred hyperfine field in the
order of one Tesla can be sufficient to induce the observed moments.

Note that no extraordinarily high Pr-Pr exchange is required to
explain the surprisingly high $T_N$ in this model, therefore the
problem of the small dependence of the transition temperature on
magnetic dilution does not occure. However, an exchange interaction of
the order of 1~K may help to understand at least qualitatively the
peculiar orientation of the Pr-moments, halfway between the
CuO$_2$-plane and the c-axis.\cite{hod221,boo306} If the Pr-moments
are induced by internal fields rather than due to $4f-4f$ exchange
interactions the magnetic structure of the Pr-sublattice is expected
to reflect the one of the Cu(2)-sublattice, a feature which might be
used to check our model. In this context it is interesting to note
that Pr orders magnetically at exceptionally high $T_N$ in a number of
related cuprates when it is situated between antiferromagnetic
(insulating) CuO$_2$-sheets.\cite{sun445,hsi444,lai442} To our
knowledge there has been only one report of coexistence of Pr-order
and superconductivity,\cite{li459} but the sample was inhomogeneous
and only partially superconducting, so magnetic order and
superconductivity might be located in different volume fractions.

\section{Conclusions}

We reported NMR/NQR investigations of the charge distribution in
Pr$_{1+x}$Ba$_{2-x}$Cu$_3$O$_{6+y}$ crystals at low temperature.
Comparison of the EFG at Cu(1) in the chains with the one in Y123
confirms the assumption of the hybridization model, that the hole states
doped with increasing $y$ into the structure are not localized in
the chain layer but in the CuO$_2$-Pr-CuO$_2$-trilayer. The relative
intensities of the resonances corresponding to the different oxygen
coordinations of Cu(1) in the chains show that there is no tendency to
form O-Cu(1)-O-chains with Pr on the RE-site, a result which fits nicely
into the trend of the interaction of oxygen on chain sites with the
RE-radius.

Our investigations of the Cu(1)- and Pr-resonance in
off-stoichiometric crystals in the Pr/Ba solid solution system show
that the accessible concentration range for $x$ in
Pr$_{1+x}$Ba$_{2-x}$Cu$_3$O$_{6+y}$ can be extended to negative $x$ by
preparation at reduced partial oxygen pressure. The fingerprint of the
$[Ba]_{Pr}$-defect in the Ba-rich crystal is a wipe-out of the
Pr-signal and an inhomogeneous internal magnetic field at the
Cu(1)-site, which we ascribe to localized singlet states in the
antiferromagnetic CuO$_2$-layers. $[Pr]_{Ba}$ in the Pr-rich crystal,
on the other hand, does not influence the Pr-resonance or induce
disorder in the magnetic Cu(2)-structure, but the structural and
charge disorder in the chain layer is evident from the inhomogeneity
of the Cu(1)-quadrupole splittings. These findings give strong support
to our earlier assignment of the Pr-signal to regular (RE-)
sites.

The Pr-resonance was observed in all crystals except one fully reduced,
stoichiometric specimen.  In order to give a consistent interpretation
of this NMR-signal as well as for the observations from INS and the
volume probes, we propose that the two electronic Pr configurations
introduced by Fehrenbacher and Rice correspond to two different ground
state symmetries of Pr in the RE-layer. We provide evidence that NMR
detects only the state with a hole localized in its $4f$-shell and the
$2p\pi$- orbitals of the surrounding oxygen ligands, while the
properties observed with space-integral techniques like INS are
dominated by the response from sites without such a localized hole.
The model is supported by our count of localized carriers with the
Cu(1)-resonance, the decreasing intensity of the Pr-resonance with
decreasing oxygen concentration $y$, by our detailed analysis of the
CEF of Pr, and it explains the discrepancy in the magnetic states of
Pr observed in NMR on one hand, and INS and susceptibility on the
other. Finally, we argue against the assumption of a strong, enhanced
$4f-4f$ exchange as the origin of the extraordinarily high Pr ordering
temperature and the suppression of superconductivity. As an
alternative we propose that the Pr-moments are induced by internal
fields which appear due to the reorientation transition of the
Cu(2)-moments.

\acknowledgements We appreciate fruitfull discussions with J. Appel and
N. H. Andersen. Thanks are due to J. K\"otzler for hospitality during
the NMR work and use of his SQUID-magnetometer to take the
susceptibility data. One of us (M.W.P.) greatly acknowledges support
by the Deutsche Forschungsgemeinschaft and by \"Osterreichische Fonds
zur F\"orderung der wissenschaftlichen Forschung under grant no.
P13568-PHY.


\bibliographystyle{prsty}
\bibliography{ord1ff,reviews,ohnekopi}

\newcommand{\noopsort}[1]{} \newcommand{\printfirst}[2]{#1}
  \newcommand{\singleletter}[1]{#1} \newcommand{\switchargs}[2]{#2#1}
\begin{thebibliography}{10}

\bibitem{rad216}
H.~B. Radousky, J. Mat. Res. {\bf 7},  1917  (1992).

\bibitem{Huefner}
S. H{\"u}fner, Adv. Phys. {\bf 43},  183  (1994).

\bibitem{pic333}
W.~E. Pickett, Physica C {\bf 289},  51  (1997).

\bibitem{tal438}
J.~L. Tallon {\it et~al.}, Phys. Rev. B {\bf 51},  12911  (1995).

\bibitem{zha439}
H. Zhang and H. Sato, Phys. Rev. Lett. {\bf 70},  1697  (1993).

\bibitem{neu416}
J.~J. Neumeier and M.~B. Maple, Physica C {\bf 191},  158  (1992).

\bibitem{cao420}
G. Cao, Y. Yu, and Z. Jiao, Physica C {\bf 301},  294  (1998).

\bibitem{ska331}
S. Skanthakumar {\it et~al.}, Phys. Rev. B {\bf 55},  R3406  (1997).

\bibitem{tom422}
Z. Tomkowicz, Physica C {\bf 320},  173  (1999).

\bibitem{fra428}
I. Francois {\it et~al.}, Phys. Rev. B {\bf 53},  12502  (1996).

\bibitem{wan423}
Y. Wang, H. Rushan, and Z.-B. Su, Phys. Rev. B {\bf 50},  10350  (1994).

\bibitem{uma318}
S. Uma {\it et~al.}, Phys. Rev. B {\bf 53},  6829  (1996).

\bibitem{fis413}
R.~A. Fisher {\it et~al.}, Physica C {\bf 235-240},  1749  (1994).

\bibitem{lun417}
P. Lundqvist {\it et~al.}, Physica C {\bf 269},  231  (1996).

\bibitem{hil147}
G. Hilscher {\it et~al.}, Phys. Rev. B {\bf 49},  535  (1994).

\bibitem{boo205}
A.~T. Boothroyd, S.~M. Doyle, and R. Osborn, Physica C {\bf 217},  425  (1993).

\bibitem{neh199}
K. Nehrke, M.~W. Pieper, and T. Wolf, Phys. Rev. B {\bf 53},  1  (1996).

\bibitem{neh352}
K. Nehrke and M.~W. Pieper, Phys. Rev. Lett. {\bf 76},  1936  (1996).

\bibitem{lop447}
M.~E. Lopez-Morales {\it et~al.}, Phys. Rev. B {\bf 41},  6655  (1990).

\bibitem{fin200}
J. Fink {\it et~al.}, Phys. Rev. B {\bf 42},  4823  (1990).

\bibitem{har410}
A. Hartmann, G.~J. Russell, W. Fentrup, and K.~N.~R. Taylor, Sol. St. Comm.
  {\bf 89},  77  (1994).

\bibitem{feh198}
R. Fehrenbacher and T.~M. Rice, Phys. Rev. Lett. {\bf 70},  3471  (1993).

\bibitem{lie197}
A.~I. Liechtenstein and I.~I. Mazin, Phys. Rev. Lett. {\bf 74},  1000  (1995).

\bibitem{maz357}
I.~I. Mazin and A.~I. Liechtenstein, Phys. Rev. B {\bf 57},  150  (1998).

\bibitem{kho000}
D. Khomskii, J. Supercond. {\bf 6},  69  (1993).

\bibitem{dre440}
S.-L. Drechsler, J. Malek, and H. Eschrig, Phys. Rev. B {\bf 55},  606  (1997).

\bibitem{gui232}
M. Guillaume {\it et~al.}, J. Phys. Cond. Matt. {\bf 6},  7963  (1994).

\bibitem{hu431}
Z. Hu {\it et~al.}, Phys. Rev. B {\bf 60},  1460  (1999).

\bibitem{che414}
J.~M. Chen {\it et~al.}, Phys. Rev. B {\bf 55},  14586  (1997).

\bibitem{mer358}
M. Merz {\it et~al.}, Phys. Rev. B {\bf 55},  9160  (1997).

\bibitem{mcc434}
R. McCormack, D. de~Fontaine, and G. Ceder, Phys. Rev. B {\bf 45},  12976
  (1992).

\bibitem{liu435}
D.~J. Liu, T.~L. Einstein, P.~A. Sterne, and L.~T. Wille, Phys. Rev. B {\bf
  52},  9784  (1995).

\bibitem{ali437}
A.~A. Aligia, J. Garces, and H. Bonadero, Physica C {\bf 190},  234  (1992).

\bibitem{hei3}
I. Heinmaa {\it et~al.}, Appl.Mag.Res. {\bf 3},  689+  (1992).

\bibitem{lue329}
H. L{\"u}tgemeier {\it et~al.}, Physica C {\bf 267},  191  (1996).

\bibitem{uim436}
G. Uimin and J. Rossat-Mignod, Physica C {\bf 199},  251  (1992).

\bibitem{amb429}
C. Ambrosch-Draxl, P. Blaha, and K. Schwarz, J. Phys. C {\bf 6},  2347  (1994).

\bibitem{ohn339}
T. Ohno, K. Koyama, and H. Yasuoka, Physica B {\bf 237-238},  100  (1997).

\bibitem{bak441}
O.~N. Bakharev {\it et~al.}, Appl. Magn. Res. {\bf 3},  613  (1992).

\bibitem{li52}
W.-H. Li, W. Lynn, S. Skanthakumar, and T. Clinton, Phys.Rev.B {\bf 40},  5300+
   (1989).

\bibitem{boo306}
A.~T. Boothroyd {\it et~al.}, Phys. Rev. Lett. {\bf 78},  130  (1997).

\bibitem{uma419}
S. Uma {\it et~al.}, J. Phys. C {\bf 10},  L33  (1998).

\bibitem{sta461}
U. Staub, Phys. Rev. Lett. {\bf 77},  4688  (1996).

\bibitem{coo402}
D.~W. Cooke {\it et~al.}, Hypewrfine Interactions {\bf 63},  213  (1990).

\bibitem{sun445}
A. Sundaresan {\it et~al.}, Phys. Rev. B {\bf 49},  6388  (1994).

\bibitem{hsi444}
W.~T. Hsieh {\it et~al.}, Phys. Rev. B {\bf 49},  12200  (1994).

\bibitem{lai442}
C.~C. Lai {\it et~al.}, Phys. Rev. B {\bf 50},  4092  (1994).

\bibitem{dro304}
H. Dr{\"o\ss}ler {\it et~al.}, Zeitsch. Phys. B {\bf 100},  1  (1996).

\bibitem{gua351}
W. Guan {\it et~al.}, Phys. Rev. B {\bf 49},  15993  (1994).

\bibitem{das403}
I. Das {\it et~al.}, Physica C {\bf 173},  331  (1991).

\bibitem{keb226}
A. Kebede {\it et~al.}, Phys. Rev. B {\bf 40},  4453  (1989).

\bibitem{zou353}
Z. Zou, K. Oka, T. Ito, and Y. Nishihara, Phys. Rev. Lett. {\bf 80},  1074
  (1998).

\bibitem{lus509}
M. Luszczek {\it et~al.}, Physica C {\bf 322},  57  (1999).

\bibitem{wid415}
K. Widder {\it et~al.}, Physica C {\bf 264},  11  (1996).

\bibitem{bre350}
E. Brecht {\it et~al.}, Phys. Rev. B {\bf 56},  940  (1997).

\bibitem{tag453}
M. Tagami {\it et~al.}, Physica C {\bf 250},  240  (1995).

\bibitem{par412}
M. Park, M.~J. Kramer, K.~W. Dennis, and R.~W. McCallum, Physica C {\bf 259},
  43  (1996).

\bibitem{tag330}
M. Tagami and Y. Shiohara, J. Crystal Growth {\bf 171},  409  (1997).

\bibitem{lin463}
T.~B. Lindemer {\it et~al.}, Physica C {\bf 231},  80  (1994).

\bibitem{ber510}
C. Bertrand, P. Galez, R.~E. Gladyshevskii, and J.~L. Jorda, Physica C {\bf
  321},  151  (1999).

\bibitem{gol000}
T. Wolf {\it et~al.}, Journal of Crystal Growth {\bf 96},  1010  (1989).

\bibitem{cha455}
C. Changkang {\it et~al.}, Physica C {\bf 214},  231  (1993).

\bibitem{mar336}
A.~J. Markvardsen {\it et~al.}, J. Mag. Mag. Mat. {\bf 177-181},  502  (1998).

\bibitem{kra424}
M.~J. Kramer {\it et~al.}, Phys. Rev. B {\bf 56},  5512  (1997).

\bibitem{boo433}
C.~H. Booth {\it et~al.}, Phys. Rev. B {\bf 49},  3432  (1994).

\bibitem{yan160}
H.~D. Yang, M.~W. Lin, C.~K. Chiou, and W.~H. Lee, Phys. Rev. B {\bf 46},  1176
   (1992).

\bibitem{fis457}
B. Fisher {\it et~al.}, Physica C {\bf 176},  75  (1991).

\bibitem{yos456}
J. Yoshida and T. Nagamo, Phys. Rev. B {\bf 55},  11860  (1997).

\bibitem{kra345}
M.~J. Kramer {\it et~al.}, Physica C {\bf 219},  145  (1994).

\bibitem{uma400}
S. Uma {\it et~al.}, J. Appl. Phys. {\bf 81},  4227  (1997).

\bibitem{and342}
N.~H. Andersen and G. Uimin, Phys. Rev. B {\bf 56},  10840  (1997).

\bibitem{lon227}
A. Longmore {\it et~al.}, Phys. Rev. B {\bf 53},  9382  (1996).

\bibitem{boo00}
A.~T. Boothroyd {\it et~al.}, Phys. Rev. B {\bf 60},  1400  (1999).

\bibitem{ros39}
N. Rosov {\it et~al.}, Physica C {\bf 204},  171  (1992).

\bibitem{gre354}
B. Grevin, Y. Berthier, G. Collin, and P. Mendels, Phys. Rev. Lett. {\bf 80},
  2405  (1998).

\bibitem{zou328}
Z. Zou, K. Oka, T. Ito, and Y. Nishihara, Jpn. J. Appl. Phys. {\bf 36},  L18
  (1997).

\bibitem{bla355}
H.~A. Blackstead {\it et~al.}, Phys. Lett. A {\bf 207},  109  (1995).

\bibitem{nar426}
V.~N. Narozhnyi {\it et~al.}, Int. J. Mod. Phys. B {\bf Proc 3SC-2 Conference,
  Las Vegas, May 1999},  1  (1999).

\bibitem{moo408}
A.~A. Moolenaar {\it et~al.}, Hyperfine Interactions {\bf 93},  1717  (1994).

\bibitem{nar425}
V.~N. Narozhnyi {\it et~al.}, Physica C {\bf 312},  233  (1999).

\bibitem{nehrkediss}
K. Nehrke, Ph.D. thesis, Universit{\"a}t Hamburg, 1996.

\bibitem{sin135}
D.~J. Singh, K. Schwarz, and P. Blaha, Phys. Rev. B {\bf 46},  5849  (1992).

\bibitem{schmenn}
S. Schmenn, {\em Untersuchung der Magnetischen Ordnung in
  Yttrium-Barium-Kupfer-Oxid und verwandten Verbindungen mittels Kernresonanz}
  (Dissertation, Juelich, 1996), (in german).

\bibitem{all349}
P. Allenspach {\it et~al.}, Zeitsch. Phys. B {\bf 95},  301  (1994).

\bibitem{tro319}
A. Trokiner {\it et~al.}, Physica C {\bf 226},  43  (1994).

\bibitem{Bleaney}
B. Bleaney,  in {\em Magnetic Properties of Rare Earth Metals}, edited by R.~J.
  Elliott (Plenum Press, New York, 1972), p.\ 383.

\bibitem{McCausland}
M.~A.~H. McCausland and I.~S. Mackenzie, Adv. Phys. {\bf 28},  305  (1979).

\bibitem{Hutchings}
M.~T. Hutchings,  in {\em Solid State Physics}, edited by F. Seitz and D.
  Turnbull (Academic Press, New York, 1964), Vol.~16, p.\ 227.

\bibitem{kas220}
A.~J. Kassman, J. Chem. Phys. {\bf 53},  4118  (1970).

\bibitem{lea000}
K.~R. Lea, M.~J.~M. Leask, and W.~P. Wolf, J. Phys. Chem. Solids {\bf 23},
  1381  (1962).

\bibitem{hod221}
J.~A. Hodges {\it et~al.}, Physica C {\bf 218},  283  (1993).

\bibitem{li459}
W.-H. Li {\it et~al.}, Physica B {\bf 206-207},  753  (1995).

\end{thebibliography}

%

%
%

\end{document}